\documentclass[11pt]{article}
\usepackage[utf8]{inputenc}
\usepackage{booktabs}
\usepackage{lscape}
\usepackage[top=2.5cm, bottom=2.5cm, left=3.5cm, right=3.5cm]{geometry}
\usepackage[T1]{fontenc}
\usepackage{graphicx}
\usepackage{blindtext}
\usepackage{comment}
\usepackage{setspace} 
\usepackage{caption}
\usepackage{chngcntr}
\usepackage[round]{natbib}
\usepackage[english]{babel}
\usepackage{amsmath, amsfonts, amssymb}
\usepackage{supertabular}
\usepackage{longtable}

\usepackage{subcaption}
\usepackage{scrextend}

\usepackage{threeparttable}
\usepackage{makecell} 
\usepackage{multirow}
\usepackage{color}
\definecolor{red}{rgb}{1,0,0}
\definecolor{blue}{rgb}{0,0,1}
\usepackage{todonotes}
\usepackage{supertabular} 
 
\begin{document}

\definecolor{colari}{rgb}{0, 0.5, 0.7}
\definecolor{coldam}{rgb}{0.1, 0.5, 0.1}
\definecolor{colkof}{rgb}{0.8,0.1, 0.3}
\newcommand{\damien}[1]{\textbf{\textcolor{coldam}{#1}}}
\newcommand{\aristide}[1]{\textbf{\textcolor{colari}{#1}}}
\newcommand{\bernard}[1]{\textbf{\textcolor{colkof}{#1}}}
\title{
Healthcare Quality by Specialists under a Mixed Compensation System: an Empirical Analysis\thanks{Support for this work was provided by the Industrial Alliance Research Chair on the Economics of Demographic Change. We gratefully acknowledge Marc Beltempo, Nicolas Jacquemet, Guy Lacroix, Pascal Saint-Amour, and Bruce Shearer  for their helpful comments and suggestions. We thank the \textit{Régie de l’assurance maladie du Québec} (Quebec's Health Insurance Board) for giving us access to the anonymized data used in this research. We wish to thank Marc-André Morin for outstanding research support.}}

\author{Damien Echevin\thanks{CRCHUS and Universit\'e de Sherbrooke; e-mail: damien.echevin@usherbrooke.ca},
Bernard Fortin\thanks{Universit\'e Laval, CIRANO and CRREP; e-mail: bernard.fortin@ecn.ulaval.ca} ~and
Aristide Houndetoungan\thanks{Thema, Cy Cergy Paris Université; e-mail: aristide.houndetoungan@cyu.fr}}

\date{September 2023}

\maketitle

\begin{abstract}
\singlespace

\noindent We analyze the effects of a mixed compensation (MC) scheme for specialists on the quality of their healthcare services. We exploit a reform implemented in Quebec (Canada) in 1999.  The government introduced a payment mechanism combining a \textit{per diem} with a reduced fee per clinical service. Using a large patient/physician panel dataset, we estimate a multi-state multi-spell hazard model analogous to a difference-in-differences approach. We compute quality indicators from our model. Our results suggest that the reform reduced the quality of MC specialist services measured by the risk of re-hospitalization and mortality after discharge. These effects vary across specialties.


\bigskip
\noindent \textbf{Keywords:} Mixed compensation; Quality of medical services; Hospital length-of-stay; Risk of re-hospitalization; Risk of death; Hazard model; Natural experiment.
\vspace{.05in}

\noindent \textbf{JEL codes}: I10; I12; I18; C41.

\end{abstract}

\thispagestyle{empty} \vspace{11in}

\eject \setcounter{page}{1}

\newpage

\onehalfspacing 

\section{Introduction}

Over the past few decades, healthcare expenditures have increased substantially and now represent an important and rising share of gross domestic product (GDP) in many developed countries. Several reasons may explain this phenomenon. On the demand side, the ageing of the population, income elasticity well above one, and, more recently, the COVID-19 pandemic have played key roles. On the supply side, introducing sophisticated new technologies is a major cause of rising costs.

In this context, one may wonder whether the resources from the healthcare sector could be allocated more effectively, especially those provided by physicians. In a nation such as Canada, where the public sector pays most doctors, physician spending represented 13.6\% of total healthcare spending in 2022. This corresponds to the second largest category after hospitals (Canadian Institute of Health Information, CIHI, 2022a)\nocite{CIHI2022a}. Moreover, physicians' behaviour influences a substantial part of expenditures in the healthcare sector (\textit{e.g.}, hospital resources, medication). Therefore, it is clear that the design of physician payment mechanisms that provide effective quantity and quality of medical services is a priority for healthcare policy. 

As is the case in many countries, three basic physician payment mechanisms are in place in Canada (CIHI, 2022b)\nocite{CIHI2022b}. These include fee-for-service (payment per service), capitation  (payment per beneficiary), and salary (payment per unit of time). There are also mixed compensation systems that combine at least two payment mechanisms. While the literature on the theoretical analysis of physician payment mechanisms regarding their behaviour and health outcomes is numerous  \citep[\textit{e.g.},][]{ellis1986, MaMcguire1997, McGuire2000, leger2007, chone2011, brosig-koch2017, Brekke2020,fortin2021}, the empirical evidence regarding these effects is relatively rare (\textit{e.g.}, see the recent literature review by \cite{{quinn2020}} on specialist physicians and \cite{kristensen2023} on general practitioners). The basic reason is the paucity of natural experiments, such as reforms of physician payment schemes, which can be used to analyze their effects on the quantity and quality of healthcare services.\footnote{There are also a few laboratory experiments on the effect of physician payment schemes on physicians' service provision \citep[\textit{e.g.},][]{green2014, brosig-koch2017}. While these analyses may help evaluate the causal effect of a pay system variation, they may also be subject to serious external validity problems.} This paper exploits one of these reforms that had been introduced in Quebec (Canada). 

In 1999, the Quebec government introduced a major mixed compensation (MC) reform for specialists (hence excluding general practitioners), combining fee-for-service (FFS) and salary systems. Before 1999, most specialist physicians in Quebec (92\%) received payments through an FFS system. Under the MC system, specialists receive a wage (\textit{per diem})\footnote{\textit{Per diems} are paid for blocks of 3.5 hours of work, paying 300~CAD each in 1999. Note that the value of the half \textit{per diem} rose to 372~CAD in 2023, an increase of 24\% since 1999, while medical service fees have almost doubled during the same period.} for time spent working in hospital, combined with a reduced fee per service (on average, 41\% of standard fees). During a \textit{per diem}, a specialist can perform clinical and non-clinical services such as administrative duties and teaching, which are unpaid under the FFS system. The MC system is optional, and unanimity by vote is required at the department level for adopting the system. Consequently, within the same department of an institution, all members should use the same compensation system. In 2021, the specialists' MC participation rate was 46\% (Quebec's Health Insurance Board, RAMQ, 2022).

Several empirical papers have studied the effect of this reform on the provision of medical services that are delivered by specialists who have adopted this system. \cite{dumont2008} provide an analysis of the effect of this reform on the volume of their clinical services. According to their results, specialists who adopted the MC system reduced their clinical services by 6.15\%. At the same time, these specialists increased their time spent on administrative and teaching duties by 7.9\%. \cite{fortin2021} have developed a structural model to study the labour supply and the quantity of clinical and non-clinical services that are provided by specialists who could adopt either a standard FFS or an MC system. Their results indicate that clinical services decreased by 5.2\% under the MC reform. 

While the reform reduced the quantity of healthcare services, a crucial issue is whether MC doctors substituted quality for the quantity of these services. Intuitively, the MC system may exert two opposite effects on the quality of clinical services. On the one hand, the introduction of \emph{per diems} and the reduction in the fees that are paid to MC doctors may induce them to spend more time and effort per service and to substitute quality for quantity of clinical services. For instance, doctors with a low marginal disutility of effort and who are highly altruistic, that is, whose welfare strongly depends upon their patients' utility, may be encouraged to spend more time treating their patients. They may also be inclined to provide them with more appropriate diagnostic services or preventive advice. On the other hand, the \textit{per diems} may support minimum-effort work standards provided by doctors with a high marginal disutility of effort or a low degree of altruism toward their patients. This is more likely to be the case in the absence of strong competition\footnote{Waiting lists under a user-free universal system like in Quebec are likely to limit the strength of competition between physicians as the supply of clinical services is the short side of the market.} or rigorous supervision by the head of the department.\footnote{Peer effects from diligent specialists in the same department may discourage shirking behaviour. Yet, this effect is likely to be reduced because the decision to adopt the MC system is made at the department level rather than at the doctor level.} Indeed, an MC specialist always receives the same \textit{per diem} at the hospital, and this payment is independent of the quality of care that is delivered to patients. This effect may reduce the quality of clinical services. Also, the reform may induce MC physicians to substitute nonclinical services for clinical services, as they are allowed to perform activities such as administrative tasks and teaching during the \textit{per diems}.  

Regarding the previous studies on the effect of MC on the quality of healthcare services, results by \cite{dumont2008} suggest that time per clinical service increased by 3.8\%  for specialists who adopted the MC system. Yet, this input-based measure of quality does not provide any information concerning the actual effect of the MC system upon the health of patients.\footnote{Some researchers use other input-based approaches that  include "best-care practices". A leading example is \cite{rosenthal2005}, who investigated whether or not patients with certain diagnoses received recommended tests for cancer and diabetes \citep[see also][]{li2014}. These measures have the advantage of being directly related to patient care. Time-based studies measure the amount of time that physicians spend with patients. \cite{MaMcguire1997} suggest using such time-based measures as a measure of quality. They argue that a careful diagnosis, treatment, and explanation of the required follow-up actions of the patient take time. However, time that is spent with patients can be noisy measures of quality, ignoring whether or not the time was necessary while affecting patients' health outcomes.} The output-based approach addresses this problem by focusing upon events that are correlated with the patient's health following a particular treatment. For example, \cite{cutler1995} measured mortality rates in hospitals (or within one year of discharge) and hospital re-admission rates as measures of the quality of care that was received. \cite{geweke2003} used mortality rates to measure hospital quality, controlling for non-random hospital admissions \citep[see also][for a survey]{Fischer2014}. \cite{Clemens2014} found that financial incentives affected the services that were provided to cardiology patients in the U.S., yet they found statistically insignificant impacts on patient health outcomes, as measured by mortality rates within four years of diagnosis.

To our knowledge, the study by \cite{Echevin2014}, henceforth denoted EF, is the only one that provides an empirical analysis of the effect of the MC reform using an output-based indicator of the quality of clinical services. These authors estimate a transition model between spells (or stays) in and out of the hospital, using panel data from a major teaching hospital in Quebec (Sherbrooke University Hospital Centre, CHUS). Their results suggest that the hospital length-of-stay (LOS) of patients treated in departments that opted for the MC system increased by 4.2\% (0.28 days). Yet, the reform did not affect the re-admission rate of these patients to the same MC department with the same diagnosis. Only two departments (Vascular Surgery and Rheumatology) were affected negatively (re-admission rates increased) by the reform.

The aforementioned studies suggest that the reform reduced the quantity of clinical services, but they are inconclusive regarding the quality of health services. Whether robust results regarding healthcare quality can be obtained when using more health outcomes, a more sophisticated econometric approach and when the analysis is applied to the jurisdiction (Quebec) as a whole is a critical question that requires further investigation.
 
Following EF, our study aims to analyze the effect of the MC reform on the quality of health services using an output-based approach. However, we extend the EF paper in five significant directions. First, we develop and estimate a multi-state multi-spell  (MSMS) proportional conditional hazard  model \citep[see][for a survey]{bijwaard2014} analogous to a difference-in-differences approach. The observational unit that is treated, \textit{i.e.}, that has adopted the MC system over the sample period, is the department (or specialty) of the treating physician in a hospital (\textit{e.g.}, Cardiology at the McGill University Health Centre, Glen Site). Our approach
assumes that a patient can move through three states over time: (Re)hospitalization, Home, and Death. Hospitalization is the original state, and death is the absorbing state. There are four (direct and indirect) transitions across the three states.\footnote{Home may include the patient’s residence or nursing, convalescent and long-term care institutions. Although patients can be followed in hospitals, our data do not provide information on the nature of other locations.} In our model, there are three potential quality indicators rather than a single one: the re-hospitalization rate, the death rate from the hospital, and the death rate from home. The average treatment effects on the treated (ATT) patients are measured by the effect of the reform on the conditional hazard (in log) of these three transitions. In some specifications, these ATT are computed for each selected specialty.

Second, we introduce unobserved heterogeneity that is correlated across states at the individual level \citep[see][]{bijwaard2014}. As is well known, failure to account for unobserved heterogeneity may be the source of bias in the duration dependence and makes the effect of covariates biased towards zero on the conditional hazard rate. We use a factor-loading approach \citep[\textit{e.g.}, see][]{berg1997} to allow for flexibility in the correlation between random effects (frailties) across states. 

Third, instead of using data from only a single hospital (CHUS), we use data from many hospitals (143) covering Quebec. Owing to our administrative panel data set, we introduce many hospital and specialty fixed effects to account for the endogeneity of the physician payment system choice at the department level after the reform.\footnote{An important advantage of having all hospitals in Quebec in our sample is the possibility of estimating the effects of the reform by taking general equilibrium effects into account. Indeed, within the same hospital, the increase in the duration of hospitalization in a treatment department may reduce the duration in a control department. In principle, comparing the same department or medical specialty in two different hospitals solves this problem, at least partly.} These covariates allow us to account for unobservable attributes of hospitals and specialties that affect both the quality of clinical services and departments' preferences for payment systems. Overall, the hospital fixed effects include as many as 99\% of patients per transition in some specifications. 

Fourth, some specifications control for several diagnosis fixed effects and a co-morbidity variable, \textit{i.e.}, the simultaneous presence of two or more medical conditions in a patient. By doing so, we attempt to control for the selection bias arising from the fact that the doctors in the departments that  have adopted the MC system may have a stronger interest in choosing patients with substantial health problems, given that time (\textit{per diem}) devoted to clinical services is not penalized when patients suffer from a serious condition. 

Finally, we develop an econometric framework inspired by \cite{athey_2006} nonlinear approach to simulate, for each specialty, the unconditional ATT of the reform on patients' duration (in days) in each state leading to a given transition (see Appendix A). This provides information for evaluating the effect of the reform based upon variables that may be most useful for policymakers.

Overall, and as is the case in EF, our empirical results suggest that the average risk of discharge to home is reduced (by 5.7\%) in departments that have adopted the MC system. This corresponds to an increase of 0.75 days in the LOS in hospital.
Also, according to our results and using our output-based indicators, the quality of healthcare services decreased in departments that adopted the MC system, based upon two quality indicators. First, in contrast with EF, our results indicate that the average risk of re-hospitalization (here, within 30 days, with the same diagnosis and in a similar department) increased by as much as 17.8\% for patients who were treated in MC departments. This corresponds to an average of three fewer days before being readmitted, for patients who are re-hospitalized within 30 days after discharge. The affected specialties are Pediatrics, Obstetrics \& gynecology, General Surgery, Psychiatry, and Internal Medicine. Second, the risk of death increased by 6.2\%  for patients at home within one year following discharge from an MC department. This corresponds to an average of 7.45 fewer days before death, for patients who died within one year before discharge. The affected specialties are
General Surgery, Urology, and Internal Medicine. However, the reform does not significantly affect the average death risk for hospitalized patients in an MC department. 

The rest of the paper is organized as follows. In Section 2,  we present the context. Section 3 introduces the data and presents some descriptive statistics. Section 4 presents our MSMS hazard model. Section 5 presents our main results. Section 6 provides a dynamic analysis.  Section 7 provides discussions and concludes.

\section{Context}

In Canada, public health insurance falls within the jurisdiction of provincial governments. For instance, Quebec's Health Insurance Plan is a public plan that provides all Quebecers with access to user-free health care, independent of their ability to pay. While there is no co-payment or user fee for care on the demand side, healthcare service rationing is a source of waiting lists that may be very long for seeing a doctor or for certain treatments in hospital. Under the FFS system, asymmetric information also pervades the healthcare market, given that patients rely on physicians’ expert knowledge in planning their medical care \citep{arrow1963}. A simple principal-agent model predicts that physicians might have an interest in manipulating information vis-à-vis their patients in such a way as to raise the demand for potentially "unnecessary" medical services and, therefore, to increase their income. This phenomenon is denoted “supplier-induced demand” in the literature \citep{evans1974} and may lead to the overconsumption of medical services provided by specialists in hospitals.\footnote{Yet, as noted by EF, FFS specialists’ incentives to induce demand for clinical services are likely to be reduced since waiting lists to see specialists are very long in Quebec, partly due to the absence of user-fee health care.} This potential problem partly explains why the government introduced the MC system in 1999. By introducing  \textit{per diems} independent of the volume of medical services and by reducing the fees per service that are paid to doctors, the reform did lead to a reduction in the quantity of clinical services that are provided by specialists, as mentioned above \citep[see][]{dumont2008,fortin2021}.\footnote{Services provided within private clinics are paid under the FFS system. Furthermore, the maximum number of half \emph{per diems} that a specialist in a hospital can claim for two weeks is 28, and these can only be claimed Monday to Friday between 7 AM and 5 PM. Once the maximum number of \emph{per diems} is reached, or when a physician works outside the \emph{per diem} claimable hours, he is paid on the FFS basis.} However, it is unclear which proportion of this reduction relates to unnecessary services.

A second reason why the government introduced the MC system is to compensate by \emph{per diems} specialties that are penalized under the FFS system. In particular, pediatricians are known to have relatively low productivity in terms of quantity, given that children and their families need special care, which often requires more time than single adults for the same medical service (with the same fee). As a result, a large proportion of pediatricians adopted the MC system. In 2021, their MC participation rate was as high as 74\% as compared with 46\% for all specialists (RAMQ, 2022).\footnote{Psychiatry is also a specialty that is not very productive in terms of volume of services given the average duration required per appointment.} 

The third reason why the government introduced the MC reform, which is the focus of this paper, is consecutive to the potential deleterious effect of FFS on the quality of clinical services. Setting in place high-powered incentives for seeing patients can lead to lower waiting times and improve access to health care in the short term, but FFS may cause some physicians to see patients too quickly, thereby neglecting the quality of care. This can lead to future costs if treatments must be repeated or health complications develop, which would require further medical care. 

When designing physician compensation contracts, the optimal healthcare policy must balance the tradeoff between quantity and quality. One basic issue is thus whether the MC reform allows to improve this tradeoff by increasing the quality of medical services provided by specialists. From a theoretical point of view, \cite{fortin2008} have developed a model of labour supply inspired by \cite{becker1973}, which attempts to evaluate the effect of the MC reform on the quantity and quality of healthcare services when an MC doctor is altruistic vis-à-vis his patients. Unfortunately, without imposing numerical values on various model parameters, the reform's effects are ambiguous on the quantity and quality of clinical services. As mentioned above, whether the net effect of the MC reform increases, decreases, or has no effect on the quality of services is the crucial empirical issue that we address in this paper using our econometric model.  

\section{Data}

The data set that was used in the paper is based on the administrative files of RAMQ, which is the agency responsible for paying physicians in Quebec. These files contain a record of every medical service performed (and billed to the government) by all physicians licensed to work in Quebec (and working in the public sector, a group representing about 98\% of all physicians in 2018). Our data set is constructed using patient-based sampling from 1996 to 2016, which includes periods before and after the reform date (1 October 1999). Our sample comprises all patients who were born in either April or October of an odd-numbered year and is restricted to those experiencing at least one hospital stay of one day or more during the sampling period.\footnote{This condition is needed in our model so that the data set provides information about the payment system (FFS \textit{vs} MC) of the department of the patient's treating physician during their stay in the hospital (it refers to the last department where they were hospitalized if they are out of the hospital). Note that the treating doctor is most often but not always part of the same department in which the patient is hospitalized, so it is assumed that the admitting department is that of the treating physician. This assumption, however, has no significant effect on our results.} There are 320,441 individuals in our sample. For each stay in hospital, we observe the date of the patient's arrival,\footnote{The first patient stay after 1 January 1996 is assumed to be an admission. This could imply a misclassification of stays in January 2016 if the patient had stayed in December 1995. However, the influence of this misclassification on the results is likely to be negligible, given the quantity of available data.} their treating physician's specialty, diagnosis, and LOS before discharge to home or death.  Moreover, data contains information on the patient's personal characteristics (age, sex and date of death, when it applies). 

\begin{figure}[!htbp]
    \begin{center}
 \caption{Proportion of Patients' Stays under Mixed Compensation}     
    \includegraphics[scale = 0.7]{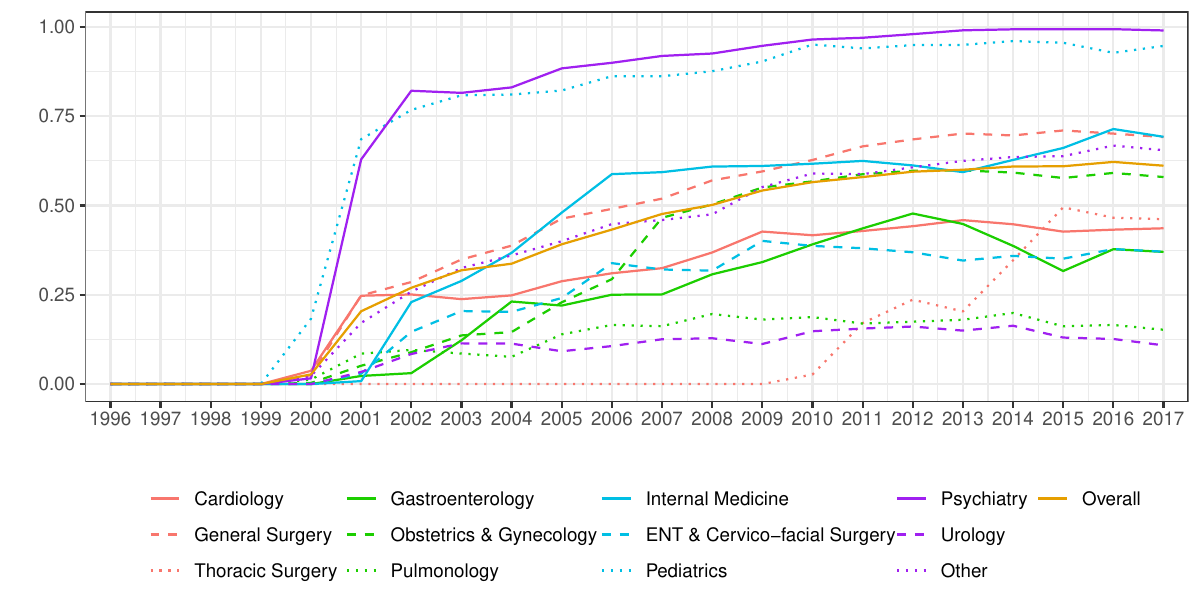}
    \label{fig:MC}
\end{center}
\scriptsize
Note: The adoption rate of mixed compensation is calculated on January 1 of each year. The reform was launched on 1 October 1999.\\
Source: RAMQ and authors' computations.
\end{figure}

We match the patient data to information concerning the payment system of the treating physician's department. Recall that most departments were paid under FFS before the reform, as shown in Figure \ref{fig:MC}, but not all of those who opted for MC adopted it as soon as the reform was implemented for the following reason.
As mentioned earlier, unanimity by vote was required for the department to adopt MC. Besides, no department returned to FFS after adopting MC during the sample period. Figure \ref{fig:MC} shows a great disparity between medical specialties. As expected, following our discussion above, Pediatrics and Psychiatry are the specialties that adopted the MC scheme the most rapidly and for which its diffusion was by far the fastest. Other medical specialties such as General Surgery, Internal Medicine, and Obstetrics \& Gynecology are average or slightly above average, while medical specialties such as Urology or Gastroenterology have much lower adoption rates. The endogeneity in the adoption of the MC scheme is discussed below.\footnote{Note that in Figure \ref{fig:MC}, the decrease in the proportion of stays of patients whose treating physician is remunerated according to the MC scheme does not indicate that a lower proportion of departments are under MC. Rather, this response indicates that a lower proportion of patients are hospitalized in these departments.}

The information regarding the patient's transitions over the sample period starts with their first admission to hospital after the beginning of the sampling period. As mentioned above, three potential states may occur after the first spell or stay in hospital: home, re-admission, and death. The data are structured to reflect these three states and the four transitions among the latter. Re-admission is strictly defined as hospitalization in the same department with the same Diagnostic-Related-Group (DRG) and not for a simple follow-up visit,\footnote{The DRG system classifies hospital cases that are expected to have similar hospital resource use into 499 groups, where DRGs are assigned by an algorithm based on the International Statistical Classification of Diseases and Related Health Problems (ICD) diagnosis codes, Current Procedural Terminology (CPT) codes, age, sex, and the presence of complications or co-morbidities. An example might be the group of females who are aged 55 and older with a Breast Cancer diagnosis, a Mastectomy procedure code, and an osteoporosis diagnosis (co-morbidity). } provided that the latter occurred within the last 30 days. After 30 days at home, the state “re-admission” is considered as right-censored. It is considered death at home if it occurs within 365 days following the previous stay in hospital. After 365 days at home, the state “death” is considered as right-censored.  Also, the state “home” is censored when the patient at home is no longer observed in the sample period. Note that when a patient is readmitted to hospital after 30 days, we consider this hospitalization a new admission. However, our model allows us to link this admission to any former hospitalization through unobserved patient random effects.

Table \ref{transitions} indicates that, on average, during our sample period, patients who were admitted to hospital have a 97.5\%  probability of being discharged to home and 2.5\% of dying in hospital. On average, patients at home have a 2\% probability of being readmitted to hospital (within 30 days after the previous stay in hospital in the same department and DRG) and 3.1\% probability of dying at home (within 365 days) and a 94.9\% probability of being censored after 365 days. 

Table \ref{descriptive} provides descriptive statistics over our sample period broken down according to each of the four transitions and censored at home. We observe 691,360 stays in hospital when the patient is discharged to home, and 17,401 stays in hospital when they die in hospital. Also, there are 14,130 stays at home with re-admission and 21,046 stays with death at home. The average LOS in hospital when the patients are discharged to home is 8.02 days, while it is 10.80 days when they die in hospital.

The average length at home when the patients are readmitted to hospital, is 13.77 days, while it is 107.23 days when they die at home. While 39\% of patients who were admitted to hospital and left for home had an MC treating specialist, 45\% of patients at home who were readmitted to hospital had an MC treating specialist during their former stay at hospital. In our sample, 57\% of patients are female, and the average age of patients is 49.3 years. We also provide information about the average Charlson Co-morbidity Index \citep{charlson1987} in each and overall transition.\footnote{The Charlson Co-morbidity Index (CCI) predicts the risk of death within one year of hospitalization for patients with specific comorbid conditions. Nineteen conditions are included in the index. Each condition was assigned a weight from 1 to 6, based on the estimated 1-year mortality hazard ratio from a Cox proportional hazard model. These weights are summed to produce the CCI score.} The average overall value of the CCI is 1.12 but reaches a value greater than 4 in transitions that end with death.

\begin{table}[htbp] 
\centering
\caption{Transition Average Probabilities}
\begin{tabular}{lcccc}   
\toprule \label{transitions}
& Home  & Re-admission & Death & Censored  \\ \midrule
Hospital & 0.975 &             & 0.025 &    0.000     \\ 
Home     &       & 0.020       & 0.031 & 0.949   \\ \bottomrule
\scriptsize
Source: RAMQ and authors' computations.

\end{tabular} 

\end{table}

\section{Estimation Strategy}

Our econometric approach attempts to take into account three basic issues. First, our model must be intrinsically dynamic to allow the patient to move to various states over time. Our model focuses upon three states: hospital, home, and death. For this purpose, we develop a recurrent multi-state multi-spell (MSMS) proportional conditional hazard model with correlated unobserved heterogeneity.\footnote{See \cite{cameron2005} ch.17-19 and \cite{bijwaard2014} for a detailed discussion of MSMS models with unobserved heterogeneity.} 

\begin{table}[htbp]
\caption{Descriptive statistics}
\centering
\footnotesize
\begin{tabular}{llrrrrr}
\toprule \label{descriptive}
&     & Ad(RAd)& Ad(RAd) & Home& Home & Censored \\
\multicolumn{2}{l}{Statistics}      &to Home & to Death & to RAd &to Death &at Home \\
\midrule
Length  & Mean & 8.02                        & 10.80                        & 13.77      & 107.23                    & 299.92                      \\
of Stay                                & SD  & 19.17                       & 25.69                        & 8.50       & 105.06                    & 121.62                      \\
                                & Min  & 1                           & 1                            & 1          & 1                         & 1                           \\
                                & Max  & 4,028                       & 1,147                        & 30         & 365                       & 365                         \\ \midrule
Mixed    & Mean & 0.39                        & 0.35                         & 0.45       & 0.34                      & 0.39                        \\
Comp.     & SD   & 0.49                        & 0.48                         & 0.50       & 0.47                      & 0.49                        \\
                                & Min  & 0                           & 0                            & 0          & 0                         & 0                           \\
                                & Max  & 1                           & 1                            & 1          & 1                         & 1                           \\\midrule
Treatment      & Mean & 0.62                        & 0.54                         & 0.67       & 0.52                      & 0.62                        \\
                                & SD   & 0.49                        & 0.50                         & 0.47       & 0.50                      & 0.49                        \\
                                & Min  & 0                           & 0                            & 0          & 0                         & 0                           \\
                                & Max  & 1                           & 1                            & 1          & 1                         & 1                           \\\midrule
Post                  & Mean   & 0.81                        & 0.82                         & 0.78      & 0.83                      & 0.81                        \\
         Reform             & SD     & 0.39                        & 0.38                         & 0.42       & 0.37                      & 0.39                        \\
                      & Min    & 0                           & 0                            & 0          & 0                         & 0                           \\
                      & Max    & 1                           & 1                            & 1          & 1                         & 1                           \\\midrule
Female         & Mean & 0.57                        & 0.47                         & 0.52       & 0.47                      & 0.58                        \\
                                & SD   & 0.49                        & 0.50                         & 0.50       & 0.50                      & 0.49                        \\
                                & Min  & 0                           & 0                            & 0          & 0                         & 0                           \\
                                & Max  & 1                           & 1                            & 1          & 1                         & 1                           \\\midrule
Age            & Mean & 48.76                       & 72.70                        & 46.44      & 72.54                     & 48.86                       \\
                                & SD   & 24.03                       & 14.90                        & 24.75      & 14.50                     & 23.83                       \\
                                & Min  & 0                           & 0                            & 0          & 0                         & 0                           \\
                                & Max  & 109                         & 108                          & 107        & 109                       & 108                         \\\midrule
Co-morbidity            & Mean & 1.05                        & 4.10                         & 1.76       & 4.21                      & 0.93                        \\
Index (CCI)                                & SD   & 1.93                        & 3.10                         & 2.51       & 3.27                      & 1.76                        \\
                                & Min  & 0                           & 0                            & 0          & 0                         & 0                           \\
                                & Max  & 18                          & 16                           & 18         & 16                        & 17                          \\\midrule
\midrule
\multicolumn{2}{l}{Number of stays}              & 691,360                     & 17,401                       & 14,130     & 21,046                    & 656,184                     \\
\multicolumn{2}{l}{Number of patients}      & 317,279                     & 17,401                       & 9,694      & 21,046                    & 311,083                     \\
\multicolumn{2}{l}{Stay$/$Patient}     & 2.18                        & 1.00                         & 1.46       & 1.00                      & 2.11             \\ \bottomrule      
\scriptsize
Source: RAMQ and authors' computations.
\end{tabular}
\end{table}

Second, we incorporate a difference-in-differences (DiD) approach in our model since one observes two basic groups in our data: the treated group of departments that are under MC at least once during the sample period; and the control group of departments that are always under FFS.

Third, since the decision of a department to adopt the MC system is optional, one must consider potential endogeneity problems in our analysis. For instance, doctors in MC departments may have a greater incentive to choose patients with complex health problems, which require more time but fewer clinical services. Also, when doctors in a department prefer activities other than clinical services when working (\textit{e.g.}, teaching, administrative tasks, leisure at work), they may have an incentive to opt for MC, in particular when the level of their \textit{per diem} is high enough to increase their income even if their fee per service is reduced.\footnote{The potential financial gain of choosing the MC scheme has been greatly reduced over time. It even became negative for most MC departments. Yet, the unanimity rule makes it difficult for physicians to return to an FFS contract that was effective almost two decades previously. In fact, no MC department has gone back to the FFS scheme in our data set.} Let us now analyze in greater detail how our econometric approach accounts for each of these three potential sources of bias.\\

\subsection{A multiple-state  multiple-spell  hazard  model}

\begin{figure}[htbp]
      \centering
  \caption{States and transitions} 
                \begin{tikzpicture}[scale=0.7]
            \centering
            \tikzstyle{quadri}=[rectangle,draw,fill=gray!20,text=blue, thick,minimum height=0.3cm]
            \tikzstyle{tranD}=[->,>=latex, color=black!30!red, thick]
            \tikzstyle{tranH}=[->,>=latex, color=black!60!green, thick]
            \tikzstyle{tranA}=[->,>=latex, color=black!10!orange, thick]
            \node[quadri, label=0:(Original State)] (A) at (0,0) {Admission (Ad)};
            \node[quadri] (H) at (-5,-3) {Home (H)};
            \node[quadri] (R) at (0,-6) {Re-admission (RAd)};
            \node[quadri, label=0:] (D) at (5,-3) {Death (D)};
            \draw[tranH] ([xshift=-2cm]A)--node [midway,above] {(1)}(H);
            \draw[tranD] (A)-- node [midway,above] {(2)}(D);
            \draw[tranA] ([xshift=2cm]H)--node [midway,above] {(3)}(R);
            \draw[tranD] (H)--node [midway,above] {(4)}(D);
            \draw[tranD] (R)--node [midway,above] {(2)}(D);
            \draw[tranH] ([xshift=-2cm]R)--node [midway,below] {(1)}(H);
        \end{tikzpicture}
\label{fig:Trans}
\end{figure}
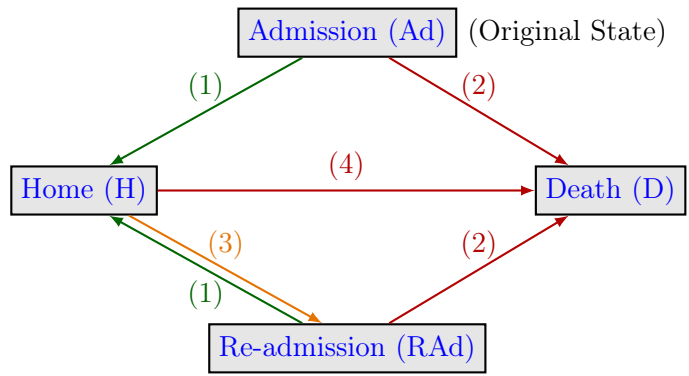

Owing to our administrative data, we assume four possible states for an individual $i$ and four transitions (see Figure \ref{fig:Trans}).\footnote{For notational simplicity, we ignore the subscript $i$ in most equations.} The original state is the first Admission to hospital (Ad) after 1 January 1996. The three other states are: Re-admission to hospital (RAd)\footnote{Re-admission is defined as hospitalization under the same DRG in the same department as the previous stay, provided that the latter occurred within the last 30 days.},  Home (H), and Death (D)\footnote{Is considered as death at home if the death occurs within 365 days of the previous stay. After 365 days at home, we assume that the individual's state is right-censored.}, the latter being the absorbing state from which the individual cannot exit. The model is in quasi-continuous time (\textit{i.e.}, the observation period is the day). Our model has four transitions: the transition from Admission to Home, the transition from Admission to Death, the transition from Home to Re-admission, and the transition from Home to Death.\footnote{The transition from Home to Admission is not considered in the modelling. This can occur in the data when it is the first admission for the patient, when the DRG or department of the new admission is different from that of the previous admission, when it is a simple control visit, or when the length of stay at home before re-admission exceeds 30 days. As we cannot consider that the patient is readmitted as a result of a previous admission and involuntarily (not for a simple follow-up visit), it is not useful to model the effect of MC on the quality of care in considering this transition.} For the sake of simplicity, we do not distinguish $Ad \to D$ from $RAd \to D$, and $Ad \to H$ from $RAd \to H$.\footnote{Our recurrent MSMS hazard model allows us to ignore the initial conditions problem.} In other words, we assume that the likelihood of being discharged or dying, conditional on the control variables, is independent regardless of whether hospitalization is the first admission or a re-admission. We make this assumption for three reasons. First, as mentioned earlier, what we consider a first admission may actually be a re-admission if the patient was admitted for the same DRG in the same department within 30 days before 1 January 1996. Second, and despite its simplicity, our most flexible econometric model has about 800 coefficients to estimate. Distinguishing these transitions would require further parameters to be estimated. Third, we control for the patient's CCI, which is assumed to reflect variations due to multiple hospitalizations for the same diagnosis if the patient's health condition becomes more critical. Therefore, if the re-admission probability of dying in the hospital is higher because the patient's health condition is more critical, we could probably control for this by considering their respective CCIs. Table \ref{Specialty} provides the list of specialties (33) and the number of patients' stays in each of their four transitions during the sample period. We define:\\

\begin{itemize}

\item $\mathcal{IS} = \{\text{Ad}, ~\text{RAd}, ~\text{H}\}$, the set of possible initial states for a transition.
\item $\mathcal{T} = \{(1),~(2), ~ (3), ~(4)\}$ the set of transitions, and
\item $\mathcal{T}(j)$ the set of transitions starting by the state $j\in\mathcal{IS}$; \textit{i.e.}, $\mathcal{T}(\text{Ad}) = \mathcal{T}(\text{RAd}) = \{(1),~ (2)\}$  and $\mathcal{T}(\text{H}) = \{(3),~ (4)\}$.

\end{itemize}

\vspace{0.3cm}


\begin{table}[htbp]
  \begin{center}
\addtolength{\leftskip} {-1.2cm}
  \caption{Transitions by specialty (1996-2016)}
    \begin{tabular}{lcccccc}
    \toprule \label{Specialty}
    \multirow{3}[2]{*}{Specialty} & Stays & Stays & \multicolumn{1}{c}{Ad(RAd)} & \multicolumn{1}{c}{Ad(RAd)} & Home  & Home \\
          & under MC & (\%)  & \multicolumn{1}{c}{to} & \multicolumn{1}{c}{to} & to    & to \\
          & \multicolumn{1}{c}{(\%)} &    & \multicolumn{1}{c}{Home} & \multicolumn{1}{c}{Death} & \multicolumn{1}{c}{RAd} & \multicolumn{1}{c}{Death} \\
    \midrule
    Allergy \& Clinical Immunology & 0.00  & 0.01  & 86    & 9     & 1     & 7 \\
    Pathology & 0.00  & 0.00  & 2     & 0     & 0     & 1 \\
    Anesthesiology & 0.15  & 0.07  & 354   & 122   & 7     & 18 \\
    Medical Microbiology \& Infectiology & 0.00  & 0.21  & 1,422  & 33    & 25    & 36 \\
    Medical biochemistry & 0.00  & 0.00  & 3     & 0     & 0     & 1 \\
    Cardiology* & 7.07  & 9.31  & 63,750 & 2,218  & 971   & 2,009 \\
    General Surgery* & 17.44 & 15.84 & 110,403 & 1,832  & 1,022  & 3,119 \\
    Orthopedic Surgery & 9.37  & 9.11  & 63,870 & 689   & 296   & 1,666 \\
    Plastic Surgery & 0.38  & 1.03  & 7,236  & 29    & 26    & 57 \\
    Thoracic Surgery* & 1.02  & 3.33  & 22,892 & 726   & 140   & 767 \\
    Dermatology & 0.01  & 0.02  & 131   & 3     & 4     & 2 \\
    Gastroenterology* & 1.21  & 2.14  & 14,580 & 570   & 465   & 768 \\
    Obstetrics \& Gynecology* & 16.41 & 18.85 & 133,454 & 136   & 945   & 367 \\
    Hematology & 1.45  & 2.07  & 13,332 & 1,355  & 1,849  & 1,773 \\
    Pulmonology* & 0.96  & 3.10  & 20,013 & 1,959  & 896   & 1,967 \\
    Internal Medicine* & 7.12  & 6.65  & 42,930 & 4,205  & 1,139  & 3,072 \\
    Physical Medicine \& Rehabilitation & 0.23  & 0.12  & 858   & 4     & 5     & 15 \\
    Neuro-Surgery & 4.23  & 2.00  & 13,743 & 464   & 207   & 724 \\
    Neuro-Psychiatry & 0.00  & 0.00  & 2     & 7     & 0     & 2 \\
    Neurology & 1.16  & 1.73  & 11,649 & 647   & 185   & 625 \\
    Ophthalmology & 0.18  & 0.64  & 4,556  & 3     & 81    & 91 \\
    ENT \& Cervicofacial Surgery* & 1.59  & 2.53  & 17,797 & 120   & 126   & 386 \\
    Pediatrics* & 11.91 & 6.66  & 47,115 & 98    & 1,323  & 55 \\
    Psychiatry* & 12.95 & 6.54  & 46,304 & 70    & 3,259  & 477 \\
    Diagnostic Radiology & 0.01  & 0.14  & 982   & 4     & 19    & 13 \\
    Radiation Oncology & 0.22  & 0.13  & 821   & 80    & 32    & 143 \\
    Urology* & 1.11  & 4.41  & 31,017 & 243   & 428   & 1,006 \\
    Nephrology & 0.97  & 1.26  & 8,324  & 629   & 179   & 453 \\
    Endocrinology \& Metabolism & 0.34  & 0.31  & 2,102  & 70    & 34    & 73 \\
    Rheumatology & 0.25  & 0.15  & 1,007  & 41    & 28    & 38 \\
    Geriatrics & 1.71  & 1.24  & 8,017  & 745   & 76    & 1,008 \\
    Medical Oncology & 0.39  & 0.33  & 2,099  & 256   & 362   & 265 \\
    Emergency Medicine & 0.14  & 0.08  & 509   & 34    & 0     & 42 \\
    All & 100.00  & 100.00  & 691,360   & 17,401    & 14,130     & 21,046 \\
    \bottomrule
    \end{tabular}%
\end{center}

\scriptsize
\textit{Note:} A star indicates a specialty that is represented by a dummy in our estimations (there are 11 stars), while the other specialties are regrouped under one dummy code (Others). Source: RAMQ and authors' computations.
\end{table}%

In our recurrent MSMS model, an individual may experience several states and the same state more than once. For example, individuals may be readmitted to hospital, go home, or readmitted again to hospital before dying (the absorbing state). The instantaneous hazard of a transition $r \in \mathcal{T}(j)$  is the instantaneous probability (or density) of leaving the state $j$ through the transition $r$, conditional on survival to time $t$. It may be conditional on some exogenous explanatory variables, such as sex, age, diagnosis, illness, co-morbidity, \textit{etc.} and unobserved heterogeneity $\nu_{(r)}$. The vector of all explanatory variables affecting the instantaneous hazard of transition $r$ is given by $\mathbf{x}_{(r)}(T)$, where $T$ is the clock time (in days) on which the patient is leaving state j. In our model, we assume that explanatory variables are not supposed to depend upon the duration in $j$, but on the clock time on which the patient is leaving state $j$ (sex, co-morbidity, age, trend, annual dummies, ...) and the transition.\footnote{Our database does not provide information on the socio-demographic characteristics of physicians, except the payment system in the department where they are attached.}

The MSMS function may also depend upon a time-invariant stochastic component that is specific to the individual and is either nonparametric or has a multivariate parametric distribution, $\boldsymbol{\nu}=(\nu_1, \nu_2, \ldots, \nu_L)$, where $L$ is the number of transitions ($\textit{i.e.}, L=4$). The distribution may also allow for interdependence between transitions. Our model is often called a mixture model, combining several random variables (including the duration of spells in each state) with different density functions.\\

\subsubsection{The multi-state multi-spell proportional conditional hazard function\label{subsubsec:hazard}} \bigskip

In our analysis, the MSMS proportional hazard of a transition $r\in \mathcal{T}(j)$  conditional on $\mathbf{x}_{(r)}(T)$ and $\nu_{(r)}$, can be written as:

\begin{equation} \lambda_{(r)}(t|\mathbf{x}_{(r)}(T),\nu_{(r)})=\lambda_{(r)0}(t)~\exp(\mathbf{x'}_{(r)}(T)\boldsymbol{\beta}_{(r)})~\nu_{(r)},\label{risque} \end{equation} 

\noindent where $\boldsymbol{\beta}_{(r)}$ is the vector (to be estimated) of the parameters of the explanatory variables for the conditional hazard of the transition $r$. \\ 

Thus, equation (\ref{risque}) includes three multiplicative regression factors that explain the conditional hazard that individual $i$ in state $j$ will go through the transition $r \in \mathcal{T}{(j)}$. 
We examine each of these three factors.

\bigskip

\noindent \textbf{The baseline hazard: $\lambda_{(r)0}(t)$} \bigskip

The hazard of the transition r (\textit{e.g.}, transiting from home to hospital) may depend upon the time that has elapsed in that initial state $j$ (\textit{e.g.}, the number of days at home). In state $j$, this dependence is captured by the baseline hazard, \textit{i.e.},\ $\lambda_{(r)0}(t)$. In the proportional hazard model, this is common to all individuals and only depends on $t$. This element is often modelled parametrically in the literature, with the distribution being assumed to belong to a given family of parametric distributions. To introduce greater flexibility, and following \cite{han1990}  and \cite{meyer1990}, we propose an estimator which assumes that the baseline hazard is piecewise  constant by time interval, so that $\lambda_{(r)0}(t, \boldsymbol{\alpha}_{(r)})$ is a step function, where $\boldsymbol{\alpha}_{(r)}$ is a vector of parameters to be estimated (see Table \ref{table_piecewise} for more details).\footnote{We also estimate the model using a Weibull function, but goodness-of-fit tests suggest that using a piecewise constant function is more appropriate.}

\bigskip 

\noindent \textbf{The regression function: $\exp(\mathbf{x'}_{(r)}(T)\boldsymbol{\beta}_{(r)})$} \bigskip

The hazard that is associated with going through transition $r$ may also depend upon some covariates other than time in state $j$. Therefore, the second element of the MSMS hazard model is a regression function that captures the effect of these variables. Since the probability of moving from one state to another is positive, we use an exponential function. The vector of explanatory variables is given by $\mathbf{x'}_{(r)}(T)$ and $\boldsymbol{\beta}_{(r)}$ is a vector of unknown parameters. In Section \ref{DiD}, we develop an expression for  $\mathbf{x'}_{(r)}(T)\boldsymbol{\beta}_{(r)}$,  which allows us to estimate the ATT of the reform on the log of the hazard for each patient's transition.\footnote{In our model, the covariates depend upon calendar time $T$, but not on duration $t$ in state $j$. In fact, only one variable changes with $t$: age. Yet, we use the age on leaving $j$, making $\mathbf{x'}_{(r)}(T)$ completely independent of the duration. The estimations are robust if we use age when the patient enters the state $j$.}\\

\noindent \textbf{Unobserved heterogeneity: $\nu_{(r)}$} \bigskip

Until now, the model ignores unobservable explanatory variables that may affect the transition risk from one state to another. This assumption is likely to be unrealistic. Indeed, the decision to leave a given state may depend upon certain factors, such as an individual's preventive health behaviour, which is usually not observed in the data set. 

An obvious example is quitting smoking to reduce the risk of mortality. To illustrate the importance of accounting for unobserved heterogeneity, one may distinguish two (unobservable) types of individuals within a cohort: those with a little penchant for ascribing to preventive health behaviour and those with a strong penchant for it. In the original state, both live at home. As time passes, the prevalence of the former type of individuals in the cohort decreases, as their risk of dying is higher relative to those who are more invested in their health. This results in a dynamic selection bias that leads, in this case, to an overestimation of the survival rate in the original state. Equivalently, the estimated hazard rate of exit (dying) decreases more with the duration spent at home than the hazard rate of a randomly selected population sample.\ 

The third element of our MSMS conditional hazard model is unobserved and time-invariant heterogeneity. Following \cite{berg1997, fougere1997, pudney2003} and \cite{lacroix2011}, we model this as a mixed weighting of the values taken by \textit{iid} random variables drawn from a standard normal distribution. The unobserved individual and transition heterogeneity are:

$$\nu_{(r)} = \exp(\omega_{{(r)}}),$$
 $$  \omega_{{(r)}} = \psi_{(r)}\varepsilon_{1} + \phi_{(r)}\varepsilon_{2},$$
where $\varepsilon_{1}$ and $\varepsilon_{2}$ are drawn independently from the standard normal distribution.
The variables $\psi_{(r)}$ and $\phi_{(r)}$ are loading factors. The correlation coefficient between transitions $r$ and $s$ can then be written as: 

\begin{equation} \label{eq:corr}Corr(\omega_{(r)},\omega_{(s)})=\frac{\psi_{(r)}\psi_{(s)}+\phi_{(r)}\phi_{(s)}}{\sqrt{(\psi_{(r)}^{2}+\phi_{(r)}^{2})(\psi_{(s)}^{2}+\phi_{(s)}^{2})}}\end{equation} From this equation, we see that our modelling of heterogeneity accounts for interdependence between the states. For identification, $\psi_{(r)} = 1$, $\forall ~r \in \{1, 2\}$ and $\phi_{(r)} = 1$, $\forall~ r \in \{3, 4\}$. Therefore, 
$$\omega_{{(r)}} = \varepsilon_{1} + \phi_{(r)}\varepsilon_{2}, ~~~\forall ~r \in \{1, 2\}$$ and $$\omega_{{(r)}} = \psi_{(r)}\varepsilon_{1} + \varepsilon_{2}, ~~~\forall ~r \in \{3, 4\}.$$

\bigskip

\subsubsection{The simulated likelihood function} \bigskip

The likelihood function of our model relies upon some basic concepts. First, the survival function, that is, the probability that the spell's duration (stay) in the original state $j$ is at least equal to $t$, can be shown to be:\footnote{For the sake of simplicity, we assume that time is continuous.} 

\begin{equation*}
    S_{j}(t|\mathbf{x}(T), \varepsilon_{1}, \varepsilon_{2}) = \prod_{s\in \mathcal{T}(j)} \exp \Big(-\exp(\mathbf{x'}_{(s)}(T)\boldsymbol{\beta}_{(s)}~ + \psi_{(s)}\varepsilon_{1} + \phi_{(s)}\varepsilon_{2})~\int_0^t\lambda_{(s)0}(\tau)~d\tau\Big).
\end{equation*}

Second, the conditional density  of duration  at time $t$, in the state $j\in\mathcal{IS}$, followed by the transition $r\in\mathcal{T}(j)$ if there is no censure in the state (that is, the likelihood of the state), is:
\begin{equation}
f_{(j)}\left(t|\mathbf{x}(T), \varepsilon_{1}, \varepsilon_{2}\right)=\left(\lambda_{(r)}(t|\mathbf{x}_{(r)}(T), \varepsilon_{1}, \varepsilon_{2})\right)^cS_{j}\left(t|\mathbf{x}(T), \varepsilon_{1}, \varepsilon_{2}\right),\label{eq:fj}
\end{equation}
where $c = 1$ if the transition is effective and $c = 0$ if the patient is censored in the state $j$. For example, after being discharged from hospital, if a patient does not die within 365 days or is not readmitted within 30 days, then we can only observe their stay in the state Home, but the patient is censored in the state. This censure is also effective though there is a new admission, but the latter is considered as readmission (maybe because the admission occurs after 30 days or because the admission is not associated with the same DRG). In this case, $c = 0$  and the likelihood of the state in Equation \eqref{eq:fj} is only the survival function.

As a patient passes through multiple states and transitions, their likelihood, which is denoted as $L_i(\boldsymbol{\theta}, \varepsilon_{1,i}, \varepsilon_{2,i})$, is the product of the likelihood in each state and for each transition, where  $\boldsymbol{\theta}$ is the vector of all parameters to be estimated. The subscript $i$ stands for the $i$-th patient.
As $\varepsilon_{1,i}$ and $\varepsilon_{2, i}$ are not observed, we have to maximize the following log-likelihood function with respect to $\boldsymbol{\theta}$:
\begin{equation}
\log~L= \sum_{i}\log\Big(\int_{\varepsilon_{1,i}}\int_{\varepsilon_{2,i}}L_i(\boldsymbol{\theta}, \varepsilon_{1,i}, \varepsilon_{2,i}) f(\varepsilon_{1,i})f(\varepsilon_{2,i})d\varepsilon_{1,i}d\varepsilon_{2,i}\Big),
\end{equation}
where $f$ is the probability density function of the standard normal distribution.

We approximate this optimization program by maximizing  the following simulated log-likelihood function with respect to $\boldsymbol{\theta}$:
\begin{equation}
\widehat{log~L}=\sum_{i}\log\Big(\sum_{m = 1}^{100}L_i(\boldsymbol{\theta}, \varepsilon_{1,i}^{(m)}, \varepsilon_{2,i}^{(m)})\Big), \label{eq:hatloL}
\end{equation}
where $\varepsilon_{1,i}^{(m)}$ and $\varepsilon_{2,i}^{(m)}$, $m = 1, \dots, 100$ are drawn from $\mathcal{N}(0, 1)$.\footnote{ Even though the literature indicates that $M$ = 20 appears adequate \citep{laroque1993}, we chose to use $M$ = 100. In general, maximization of the simulated maximum likelihood function generates convergent estimators if $\sqrt N/M $ when $N \to +\infty$ and $M \to +\infty$ \citep{gourieroux1991}, and $L_{i}$ is the contribution of individual $i$ to the total likelihood.}\\ 

\subsection{DiD estimation in the multi-state multi-spell hazard model}\label{DiD}

Our framework to estimate the MC effect can be analyzed using a linear difference-in-differences (DiD) model using a general approach inspired by \cite{athey_2006}.  The econometric object that we model is the conditional hazard $\lambda_{(r)}(t|\mathbf{x}_{(r)}(T),\nu_{(r)})$, expressed in $\log$. The set of treated units varies across time. Over a very long period, all hospital departments could adopt MC, and the set of non-treated units would then be empty. Thus, we refer to non-treated units in our sample as the departments that do not adopt MC, \textit{i.e.}, before 1 January 2017. In the absence of reform (referred to as $N$), we assume that the regression function $\mathbf{x'}_{(r)}(T)\boldsymbol{\beta}_{(r)}$ in the log of the conditional hazard function (see Section \ref{subsubsec:hazard}) is defined by 
\begin{equation}
    z_{(r)}(T)^N = h_{(r)}(T) + \underbrace{\sum_{k = 1}^{K^D}\beta_{(r)k}^{D}D_{k}(T) + \sum_{k = 1}^{K^I}\beta_{(r)k}^{I}I_{k}(T) + \mathbf{\tilde{x}'}_{(r)}(T)\boldsymbol{\tilde\beta}_{(r)}}_{U_{(r)}(T)}, \label{eq:xbN}
\end{equation}

\noindent where $h_{(r)}(T)$ controls for time effect on the conditional hazard, $D_{1}(T)$, \dots, $D_{K^D}(T)$ are $K^D$ specialty dummy variables, $I_{1}(T)$, \dots, $D_{K^D}(T)$ are $K^I$ hospital dummy variables, $\mathbf{\tilde{x}'}_{(r)}(T)$ includes other control variables, and $\beta_{(r)k}^{D}$, $\beta_{(r)k}^{I}$, $\boldsymbol{\tilde\beta}_{(r)}$ are unknown parameters. 

Here, an important remark is in order. For the sake of parsimony and to make the model tractable, we do not cross dummy variables for the specialties with those for the hospitals. This would imply a maximum of $33 \times 143 = 4719$ parameters to estimate per transition. Rather, we assume the separability of the specialty dummies for which we are interested in the MC effect and the dummies for the largest hospitals (see below).\footnote{Therefore, one limitation of our DiD approach is that it cannot include a dummy for each department, as this would make our MSMS model intractable. }

We include in $h_{(r)}(T)$ a linear combination of annual dummy variables as well as linear and quadratic quarterly trends. We also cross the quarterly trends with the specialties to take into account technological progress heterogeneity across specialties. We control for patients' observable characteristics in $\mathbf{\tilde{x}'}_{(r)}(T)$ such as age, sex, and some patients' health characteristics when they are admitted to hospital (see below).
On one hand, we include in  $\mathbf{\tilde{x}'}_{(r)}(T)$ departments' characteristics, such as socio-sanitary regions (19 dummies) and the number of specialist physicians per department. 

On the other hand, we specify the term $\mathbf{x'}_{(r)}(T)\boldsymbol{\beta}_{(r)}$ in the log of the conditional hazard function for the departments that have adopted MC (referred to as $I$) as

\begin{equation}
    z_{(r)}(T)^I = \theta_{(r)}MC(T) + \underbrace{h_{(r)}(T) + U_{(r)}(T)}_{z_{(r)}(T)^N}, \label{eq:xbI}
\end{equation}

\noindent where $MC(T) = 1$ if the department of the treating physician has adopted MC before the date $T$ and $MC(T) = 0$ otherwise. The parameter $\theta_{(r)}$ captures the effect of the MC on the expected log of the conditional hazard. Since the conditional hazard model is linear in logarithms, this coefficient also represents the ATT of the log of the hazard for the transition $r$. Most importantly, this measure neither depends upon unobserved patient heterogeneity $\nu_{(r)}$ nor upon the control variables that are included in $z_{(r)}(T)^N$ (see Appendix A). In Equation \eqref{eq:xbI}, the MC effect is constant, irrespective of the specialty and adoption date. We present the equation in this way to ease the notational burden. In some specifications, we decompose the MC effect (ATT) by specialty or by the adoption time interval to allow for dynamic analysis.

One important identification strategy, when using the DiD approach as applied to our hazard framework, is to assume that in the absence of the MC reform and given $h_{(r)}+U{(r)}(T)$, the hazard is the same for the treated and non-treated units in a given time period (parallel trend assumption). While it is impossible to test this hypothesis, one can consider a modified version of Equation \eqref{eq:xbN} before the reform, \textit{i.e.}, on the period 1996-01-01 to 1999-08-31, which is given by:

\begin{equation}
   \textstyle \tilde{z}_{(r)}(T)= \tilde{MC}\tilde{\theta}_{1(r)}{Q}(T) + \tilde{MC}\tilde{\theta}_{2(r)}Q(T)^2 + h_{(r)}(T) + U_{(r)}(T), \label{eq:txbN}
\end{equation}

\noindent where $\tilde{MC} = 1$ for the specialty of the treating physician that will adopt MC later and $\tilde{MC} = 0$ for the non-treated departments. Finally, $Q(T)$ and $Q(T)^2$ respectively represent the linear and quadratic quarterly trends. Testing the hypothesis that the hazard is the same in a given time period before the reform, regardless of whether the department will adopt MC or not, is equivalent to jointly testing $\tilde{\theta}_{1(r)} = 0$ and $\tilde{\theta}_{2(r)} = 0$ when we also control for all other covariates. The intuition underlying this test is to verify whether treated departments evolved differently from other departments before the reform. We also find similar results when we generalize the quadratic polynomial trends in equation \eqref{eq:txbN} to cubic trends.

\subsection{Endogeneity in the multiple-spell multiple-state hazard model}

Our model may be subject to serious problems of endogeneity. The basic reason is that the decision to adopt the MC system after the reform is optional at the department level. For instance, this may be the source of a selection bias, given that some variables affecting the department's decision to choose MC may also directly influence the quality of services that are delivered. Unfortunately, we do not have an instrument that affects the decision for a department to adopt MC, but that is not correlated with the random terms of patients or departments that affect patients' transitions.\footnote{See \cite{walter2015} for a paper that develops an approach to introduce an instrumental variable into a hazard model based upon a control function.} We adopt the following solutions to take this problem into account.

First, owing to our panel data set, we introduce fixed effects both for hospitals where patients have been (currently or lastly) admitted and for their training doctor's specialty. This approach has a number of advantages. It allows us to account for the unobservable characteristics of the hospital that may influence both the decision of its departments to operate under MC and the quality of healthcare services delivered. For instance, peer effects and homophily between doctors within a hospital may be such that their preferences regarding professional behaviour and choices are similar. Also, the fixed effect for the training doctor's specialty  allows us to account for the unobservable attributes of his specialty that may influence his preferences for the payment systems (\textit{e.g.}, pediatrists and psychiatrists prefer the MC system while radiologists prefer the FFS system), and his professional behaviour.

Here, an important remark is in order. We did not include dummies for all 143 hospitals that are used in our model. To render the model more parsimonious and tractable, we limit the number of dummies in some specifications to the largest 80 hospitals since the latter includes as many as 99\% of patients who have moved from one state to another. Also, we introduce 12 dummies for specialties (see  Table \ref{Specialty}).\footnote{There are 33 specialties in our database. We regroup 22 of them in the category "Others" for two reasons. First, the number of hospitalized patients is very low in our sample in some specialties, at least for some transitions (\textit{e.g.}, diagnostic radiology, neonatology, radio-oncology). Second, only a few physicians opted for MC in some specialties (e.g., microbiology–infectiology, diagnostic radiology).} These 92 dummies (which involve 368 parameters to estimate since there are four transitions) allow us to account for unobservable time-invariant characteristics of the 80 hospitals and 12 specialties, which in turn may influence their decision to adopt MC. Ideally, we should include $(80 \times 12 = 960$ dummies) for each transition to account for the unobservable characteristics of each department in each hospital. However, such a model is not estimable since it would involve at least $960\times 4= 3,840$ parameters to estimate.

Second, it is arguable that departments in which physicians prefer to treat patients with complex health problems and strong co-morbidity problems have more incentive to adopt the MC system. The basic reason is that being paid with \textit{per diems} and low fees per service under MC penalize a physician less than under FFS when their patients' treatments are time-intensive. This may be the source of a serious selection bias in identifying the reform's effect on patients' transitions. For instance, the estimated effect of MC on the risk of dying when the patient is at home is likely to be overestimated as long as MC physicians treat patients with more complex diseases than under FFS. To take this heterogeneity problem into account, we have introduced a large number of control variables. In particular, we have included 18 diagnosis dummy variables per transition as an indicator of the patient's health when admitted to hospital.\footnote{Diagnoses are grouped into 18 categories according to the three-digit sections of the International Classification of Diseases (ICD-9). These categories include infectious and parasitic diseases, neoplasms, endocrine, nutritional, and metabolic diseases, immunity disorders, diseases of the blood and blood-forming organs, mental disorders, diseases of the nervous system and sense organs, diseases of the circulatory system, diseases of the respiratory system, diseases of the digestive system, diseases of the genitourinary system, complications of pregnancy, childbirth and the puerperium, diseases of the skin and subcutaneous tissue, diseases of the musculoskeletal system and connective tissue, congenital anomalies, certain conditions originating in the perinatal period, symptoms, signs, and ill-defined conditions, injury and poisoning, external causes of injury and supplemental classification.} Also, as discussed earlier, we have introduced the Charlson Co-morbidity Index (CCI score) as an additional control variable.

Third, we have introduced a control variable for the number of specialists in the patient's department. This variable may affect the decision to adopt the MC system as it may be more difficult to reach unanimity when the department size is large. Yet, this variable may also directly influence the quality of healthcare services, given that the number of physicians who are substitutes (or complements) for the treating physician increases when the size of the patient's department increases. Note that the so-called incidental parameter problem is not an issue here since the asymptotics are on patients and not on hospitals or departments, which are assumed to be fixed in the analysis.\

Fourth, using equation (\ref{eq:txbN}), we have performed parallel trend tests based on the estimated parameters of the model when one focuses solely on the pre-reform period. We compare the evolution of the (linear and quadratic) quarterly trends in departments that would later switch to MC \textit{vs} those that would stay under FFS. This test is discussed at the end of the subsection \ref{DiD} (see Equation (\ref{eq:txbN})). 

Section \ref{results} presents a robustness analysis of our results, depending upon the introduction (or not) of various control variables to account for endogeneity (see Tables 5 and 6). It also presents the results of our parallel trend tests.

\section{Results}\label{results}

Table \ref{table_piecewise} provides the piecewise constant intervals (measured in days) used in the baseline hazard of each transition. They have been selected after several trials as a function of the relative importance of observations in each interval.\footnote{The estimated effect of the reform on each transition risk is robust to the choice of these intervals.} In the basic specifications provided in Table \ref{table_overall}, we assume that the reform's effect in MC departments on the risk of a given transition is the same for each specialty. Model 1 presents the most general specification, including unobservable heterogeneity (represented by four loading factors: two $\phi$ and two $\psi$) and all explanatory variables. Also, this model includes as many as 80 dummies for the largest hospitals. Introducing a larger number of dummies for hospitals would render the model intractable, given that it is estimated using a simulated maximum likelihood approach. \
\begin{table}[htbp]
  \centering
  \caption{Piecewise constant intervals (measured per day) for each transition}
    \begin{tabular}{lcccc}
    \toprule \label{table_piecewise}
          & Ad(RAd) → Home & Ad(RAd) → Death & Home → RAd & Home → Death \\
    \midrule
    $\alpha_1$ & [1, 2[ & [1, 2[ & [1, 3[ & [1, 4[ \\
    $\alpha_2$ & [2, 3[ & [2, 5[ & [3, 6[ & [4, 16[ \\
    $\alpha_3$ & [3, 4[ & [5, 10[ & [6, 8[ & [16, 31[ \\
    $\alpha_4$ & [4, 5[ & [10, 16[ & [8, 12[ & [31, 51[ \\
    $\alpha_5$ & [5, 6[ & [16, 29[ & [12, 16[ & [51, 81[ \\
    $\alpha_6$ & [6, 8[ & [29, Inf[ & [16, 19[ & [81, 121[ \\
    $\alpha_7$ & [8, 11[ &       & [19, 22[ & [121, 181[ \\
    $\alpha_8$ & [11, 18[ &       & [22, 26[ & [181, 261[ \\
    $\alpha_9$ & [18, Inf[ &       & [26, 30[ & [261, 365[ \\
    \bottomrule
    \end{tabular}%
\end{table}%

\begin{table}[htbp]
  \begin{center}
\addtolength{\leftskip} {-3cm}
  \caption{Effect (in log) of the reform on patients' risk of transition (Overall)}
    \resizebox{1.4\textwidth}{!}{
    \begin{tabular}{lccccccccccr}
    \toprule \label{table_overall}
          & \multicolumn{2}{c}{Model 1} & \multicolumn{2}{c}{Model 2} & \multicolumn{2}{c}{Model 3} & \multicolumn{2}{c}{Model 4} & \multicolumn{2}{c}{Model 5} & \multicolumn{1}{c}{Parallel trend} \\
    Variables & Estimate & SD    & Estimate & SD    & Estimate & SD    & Estimate & SD    & Estimate & SD    & \multicolumn{1}{c}{p-value} \\
    \midrule
    \textbf{1. Ad(RAd) → Home} &       &       &       &       &       &       &       &       &       &       &  \\
    Mixed Compensation & -0.057 & 0.005 & -0.047 & 0.003 & -0.082 & 0.004 & -0.082 & 0.004 & -0.081 & 0.004 & \multicolumn{1}{c}{0.071} \\
    Female & -0.006 & 0.005 & -0.018 & 0.003 & -0.002 & 0.005 & 0.034 & 0.005 & -0.002 & 0.005 &  \\
    Age   & -0.003 & 0.000 & -0.005 & 0.000 & -0.001 & 0.000 & 0.000 & 0.000 & -0.001 & 0.000 &  \\
    (Age/150)$^2$ & -0.021 & 0.001 & -0.009 & 0.000 & -0.022 & 0.001 & -0.032 & 0.001 & -0.022 & 0.001 &  \\
    $\phi$ & 0.001 & 0.003 & -     & -     & 0.002 & 0.003 & -0.005 & 0.003 & 0.002 & 0.003 &  \\
    \midrule
    \textbf{2. Ad(RAd) → Death} &       &       &       &       &       &       &       &       &       &       &  \\
    Mixed Compensation & -0.003 & 0.023 & 0.004 & 0.022 & 0.017 & 0.021 & 0.008 & 0.021 & 0.016 & 0.021 & \multicolumn{1}{c}{0.954} \\
    Female & -0.079 & 0.017 & -0.112 & 0.016 & -0.077 & 0.017 & -0.151 & 0.017 & -0.077 & 0.017 &  \\
    Age   & 0.023 & 0.004 & 0.017 & 0.003 & 0.024 & 0.003 & 0.053 & 0.003 & 0.024 & 0.003 &  \\
    (Age/150)$^2$ & 0.006 & 0.004 & 0.020 & 0.004 & 0.005 & 0.004 & -0.025 & 0.004 & 0.005 & 0.004 &  \\
    $\phi$ & -0.342 & 0.019 & -     & -     & -0.346 & 0.019 & -0.344 & 0.017 & -0.347 & 0.019 &  \\
    \midrule
    \textbf{3. Home → RAd} &       &       &       &       &       &       &       &       &       &       &  \\
    Mixed Compensation & 0.178 & 0.027 & 0.186 & 0.026 & 0.222 & 0.026 & 0.231 & 0.026 & 0.219 & 0.026 & \multicolumn{1}{c}{0.189} \\
    Female & 0.002 & 0.020 & 0.034 & 0.018 & 0.000 & 0.020 & -0.039 & 0.020 & 0.000 & 0.020 &  \\
    Age   & -0.009 & 0.002 & -0.007 & 0.002 & -0.015 & 0.002 & -0.012 & 0.002 & -0.015 & 0.002 &  \\
    (Age/150)$^2$ & 0.005 & 0.003 & -0.004 & 0.003 & 0.011 & 0.003 & 0.016 & 0.003 & 0.011 & 0.003 &  \\
    $\psi$ & -0.199 & 0.015 & -     & -     & -0.189 & 0.015 & -0.283 & 0.015 & -0.189 & 0.015 &  \\
    \midrule
    \textbf{4. Home → Death} &       &       &       &       &       &       &       &       &       &       &  \\
    Mixed Compensation & 0.062 & 0.022 & 0.048 & 0.020 & 0.069 & 0.021 & 0.060 & 0.021 & 0.065 & 0.021 & \multicolumn{1}{c}{0.887} \\
    Female & -0.219 & 0.017 & -0.171 & 0.014 & -0.220 & 0.017 & -0.327 & 0.017 & -0.219 & 0.017 &  \\
    Age   & 0.004 & 0.003 & 0.012 & 0.003 & 0.008 & 0.003 & 0.031 & 0.003 & 0.008 & 0.003 &  \\
    (Age/150)$^2$ & 0.053 & 0.004 & 0.035 & 0.003 & 0.049 & 0.004 & 0.036 & 0.004 & 0.049 & 0.004 &  \\
    $\psi$ & -0.608 & 0.016 & -     & -     & -0.598 & 0.016 & -0.789 & 0.016 & -0.599 & 0.016 &  \\
    \midrule
    \textit{log(likelihood)} & \multicolumn{2}{c}{-2346071} & \multicolumn{2}{c}{-2302170} & \multicolumn{2}{c}{-2350172} & \multicolumn{2}{c}{-2368791} & \multicolumn{2}{c}{-2350320} &  \\
    Number of patients & \multicolumn{2}{c}{320.441} & \multicolumn{2}{c}{320,441} & \multicolumn{2}{c}{320,441} & \multicolumn{2}{c}{320,441} & \multicolumn{2}{c}{320,441} &  \\
    Number of observations & \multicolumn{2}{c}{1,400,121} & \multicolumn{2}{c}{1,400,121} & \multicolumn{2}{c}{1,400,121} & \multicolumn{2}{c}{1,400,121} & \multicolumn{2}{c}{1,400,121} &  \\
    \midrule
    Unobserved heterogeneity & \multicolumn{2}{c}{Yes} & \multicolumn{2}{c}{No} & \multicolumn{2}{c}{Yes} & \multicolumn{2}{c}{Yes} & \multicolumn{2}{c}{Yes} &  \\
    Hospital FE & \multicolumn{2}{c}{80 largest} & \multicolumn{2}{c}{80 largest} & \multicolumn{2}{c}{10 largest} & \multicolumn{2}{c}{10 largest} & \multicolumn{2}{c}{10 largest} &  \\
    Charlson co-morbidity index & \multicolumn{2}{c}{Yes} & \multicolumn{2}{c}{Yes} & \multicolumn{2}{c}{Yes} & \multicolumn{2}{c}{No} & \multicolumn{2}{c}{Yes} &  \\
    Department size (number of specialists) & \multicolumn{2}{c}{Yes} & \multicolumn{2}{c}{Yes} & \multicolumn{2}{c}{Yes} & \multicolumn{2}{c}{Yes} & \multicolumn{2}{c}{Yes} &  \\
    12 Specialty FE & \multicolumn{2}{c}{Yes} & \multicolumn{2}{c}{Yes} & \multicolumn{2}{c}{Yes} & \multicolumn{2}{c}{Yes} & \multicolumn{2}{c}{Yes} &  \\
    18 Diagnoses FE & \multicolumn{2}{c}{Yes} & \multicolumn{2}{c}{Yes} & \multicolumn{2}{c}{Yes} & \multicolumn{2}{c}{Yes} & \multicolumn{2}{c}{Yes} &  \\
    19 Region FE & \multicolumn{2}{c}{Yes} & \multicolumn{2}{c}{Yes} & \multicolumn{2}{c}{Yes} & \multicolumn{2}{c}{Yes} & \multicolumn{2}{c}{Yes} &  \\
    Linear \& quadratic trends & \multicolumn{2}{c}{Yes} & \multicolumn{2}{c}{Yes} & \multicolumn{2}{c}{Yes} & \multicolumn{2}{c}{Yes} & \multicolumn{2}{c}{No} &  \\
    Year FE & \multicolumn{2}{c}{Yes} & \multicolumn{2}{c}{Yes} & \multicolumn{2}{c}{Yes} & \multicolumn{2}{c}{Yes} & \multicolumn{2}{c}{Yes} &  \\
    \bottomrule
    \end{tabular}%
    }
\end{center}
\scriptsize

\textit{Note:} All specifications use a multi-state multi-spell proportional conditional hazard model. The baseline hazards are piecewise constant. In all models (except Model 2), unobserved heterogeneity is modelled using a mixed weighting of the values taken by \textit{iid} random variables drawn from a standard normal distribution. The parallel trend p-values correspond to Model 3. All the models (except Model 2) are estimated using simulated Maximum Likelihood. Model 2 does not account for unobserved heterogeneity and is estimated using Maximum Likelihood.

\end{table}

Results for the first transition in model 1 suggest that in MC departments, the patient's risk of being discharged to home decreases on average by 5.7\%, compared with FFS departments (= ATT for the first transition). This means that patients whose treating doctor is under MC tend to stay longer in the hospital than when their treating doctor is under FFS, \textit{ceteris paribus}. Besides, results from the second transition in model 1 indicate that the risk that a hospitalized patient dies in hospital is not significantly influenced by the system under which his treating doctor is paid. \

Importantly, results from the third transition suggest that the risk of patient re-admission with the same diagnosis and in a similar department within thirty days following discharge increases by 17.8\%  when their treating doctor is under MC. This suggests that the MC system has a strong and negative effect on the quality of healthcare services insofar as it increases the risk of re-admission in hospital.\

Results from the fourth transition indicate that the risk of death within one year after a patient's discharge to home increases by 6.2\% when the treating doctor is under MC. 
Therefore, results from this transition would suggest that the MC reform reduced healthcare service quality. \

Model 3 provides one way to check for the presence of endogeneity in the estimates of the reform's effect on the four transitions. This model reduces the number of hospital dummies to the ten largest hospitals. This corresponds to 36\% of patients' transitions. Our results suggest that the estimated effect of being treated in an MC department increases (in absolute value) for all transitions in Model 3. The estimated negative effect increases from 0.057 to 0.082 in transition 1 and from 0.003 to 0.017 in transition 2. In the latter case, the estimated coefficients are not significant in any model. The estimated positive effect of being treated in an MC department slightly increases from 0.178 to 0.222 in transition 3 and from 0.062 to 0.069 in transition 4.  Therefore, one can conclude that endogeneity affects the MC coefficient in transition 1, but it does not seem very important in the other transitions, at least based upon the number of hospital dummies. \

Model 4 is similar to Model 3 except that an additional potential bias is introduced in the former model, given that the Charlson Co-morbidity Index has been removed from the set of explanatory variables. When one compares Model 4 to Model 3, results indicate that the estimated effects of being treated by an MC doctor on transitions seem quite robust to the absence of the co-morbidity index. The estimated effect of the reform on transition 1 does not change, while the effects of both models are not significant in transition 2. Also, the estimated effects of the reform on transitions 3 and 4 are very close.\

Model 5 is quite similar to Model 3 except that linear and quadratic quarterly trends have been removed from Model 1.\footnote{Yearly fixed effects have been introduced in all models.} Results also suggest that the reform's effect on transitions in MC departments is very robust to such a change. Another way to (informally) check for the presence of endogeneity is to run parallel trend tests using the pre-reform period for each of the four transitions, as explained above. For this purpose, we use Model 3 with the ten largest hospitals, as estimating a model with 80 hospitals, such as Model 1, is intractable when the sample is limited to the pre-reform period. Results that are associated with the p-value provided in the last column of Table  \ref{table_overall} indicate that we cannot reject the parallel test hypothesis for all transitions at the 5\% level.

Therefore, we can conclude that based on our four complementary solutions to take endogeneity into account, it is apparently not a severe problem in our analysis of the effect of reform on the quality of healthcare services by MC specialists; an exception could be on the risk of transition from hospital to home (transition 1). This latter caveat originates from the sizeable disparity between the estimated MC coefficients of Model 1 and Model 3 (quite large), where both coefficients are negative and significant. This difference could be interpreted by considering that in some smaller hospitals, the constraints on beds are greater, which could reduce the effect of the MC scheme on the duration of hospitalization while, at the same time, decreasing the incentive to opt for this contract.

Regarding the effects of control variables such as sex and age on the four transitions, the results that were obtained are consistent with our expectations. Being female increases time spent in hospital, reduces the risk of dying in hospital when hospitalized, and reduces the risk of dying at home within one year after being discharged from hospital. Likewise, being older increases the LOS in hospital at an increasing rate, yet increases the risk of dying in hospital when hospitalized, while reducing the risk of being readmitted within 30 days,\footnote{This may be partly explained by the fact that older people are more likely to leave hospital for a convalescent home or a long-care institution, for a given health problem.} with the rate increasing the risk of dying within one year after being discharged.

\begin{figure}[!t]
  \begin{center}
 \addtolength{\leftskip} {-1.5cm}  
    \caption{Hazards with and without heterogeneity (Model 1 and Model 2)}
    \includegraphics[scale = 0.7]{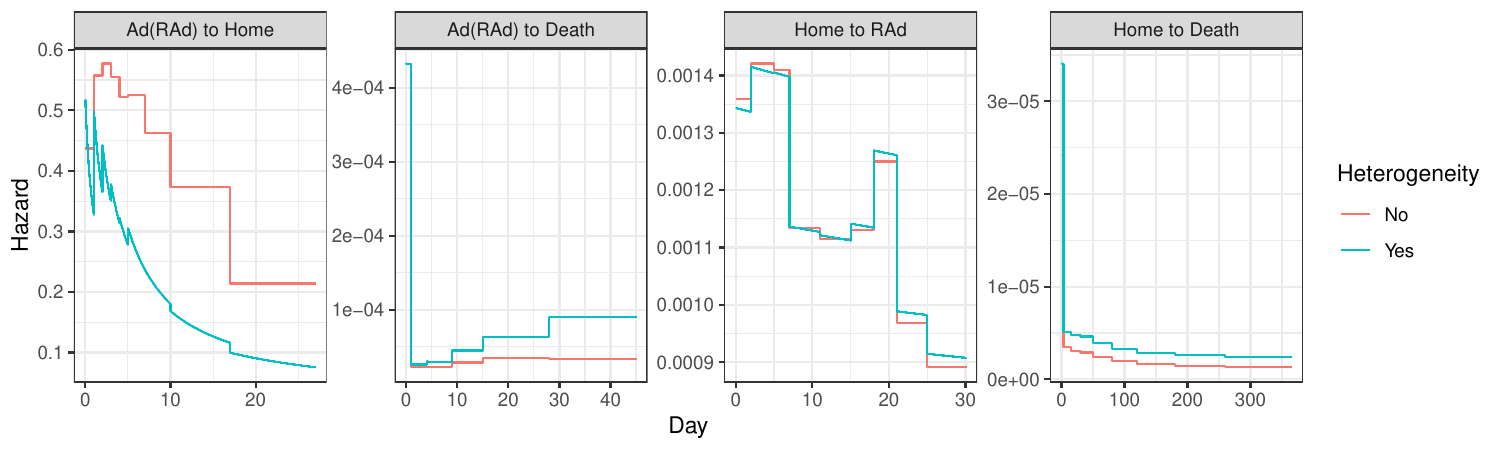}
    \label{fig:Hazard}
\scriptsize
\textit{Note:} For simplicity’s sake, all explanatory variables are fixed at zero.
\end{center}
\end{figure}

In Model 1, the loading factor parameters $\phi$ and $\psi$, which account for the (time-invariant) unobserved heterogeneity, are significant in transitions 2 to 4, but not in transition 1 that is associated with the risk (per day) of a patient being discharged. To check whether the parameters of interest are robust to this problem, we present the results of Model 2, which assumes that the $\phi$'s and $\psi$'s are zero (no unobserved heterogeneity). We see that the estimated MC coefficients do not change much when comparing Models 1 and 2. For instance, the reform's effect on the risk of being readmitted to hospital (transition 3) when the patient's treating physician is under MC is 0.178 in Model 1 and 0.186 in Model 2. The effect of MC on the risk of transition from home to death (transition 4) is 0.062 in Model 1 and 0.048 in Model 2.
Note, however, that the hazards strongly vary between these two models in some transitions. As mentioned above, the basic reason is that failure to take unobserved heterogeneity into account may introduce a bias in the relationship between the hazard and duration. Thus, in Figure \ref{fig:Hazard}, one sees that in transition 1, the hazard without heterogeneity (Model 2) is larger than the hazard with heterogeneity (Model 1) for any given level of duration in the state (positive bias).\footnote{The bias can be positive or negative, depending upon the density function of $\nu_{(r)}$. A heuristic proof of this proposition is that in our model, the hazard is given by $\lambda_{(r)0}(t)~\exp(\mathbf{x'}_{(r)}(T)\boldsymbol{\beta}_{(r)})~\nu_{(r)}$, with $\nu_{(r)} = \exp(\psi_{(r)}\varepsilon_{1} + \phi_{(r)}\varepsilon_{2})$. When heterogeneity is removed, $\nu_{(r)} = 1$. The bias is negative if $\nu > 1$ in the case of heterogeneity. If $\nu < 1$, then the bias is positive.}

We also show in Table \ref{corr} that the correlation between the various transitions is either positive or negative and statistically significant given the results that are presented in the previous tables. In particular, we observe that a longer hospital stay, \textit{ceteris paribus}, is associated with a lower risk of death in the hospital, but associated with a higher risk of re-admission and a higher risk of death outside of the hospital. This may be due to higher risks of complications following longer hospital stays, given patients' unobservable attributes. In addition, the risk of death in hospital is negatively correlated with the risks of re-admission and death outside of hospital. Therefore, if a patient lasts in hospital before dying, even though they would not have died there, they are at greater risk of being readmitted or dying at home. Finally, the risk of re-admission is positively correlated with the risk of death at home, which is expected, given the unobservable patient characteristics that can lead both to re-hospitalization and to death.

\begin{table}[htbp]
  \begin{center} 
 \addtolength{\leftskip} {-0.5cm}  
  \caption{Correlation between the transitions}
  \small 
\begin{tabular}{lcccc}
\toprule \label{corr}               & Ad(RAd) → Home & Ad(RAd) → Death & Home → RAd & Home → Death \\
               \midrule
Ad(RAd) → Home      & 1.00      & 0.95       & -0.19          & -0.52        \\
Ad(RAd) → Death     & .      & 1.00       & -0.50          & -0.77        \\
Home → RAd & .     & .      & 1.00           & 0.94         \\
Home → Death   & .     & .      & .           & 1.00        \\ \bottomrule
\end{tabular}
\end{center}
\scriptsize
\textit{Note:} This table computes $Corr(\omega_{(r)},\omega_{(s)})$, the correlation between the unobserved heterogeneity of the transition (r) and that of the transition (s) (see Equation \eqref{eq:corr}). The transition (r) is indicated in the first row, and the transition (s) in the first column. A positive correlation means that a higher duration in the transition (r) is associated with a lower probability for the transition (s).
\end{table}

Table \ref{table_speciality} provides results using the same five models as in Table \ref{table_overall}  but allows the MC reform effect (= ATT) to vary across specialties. As noted above, we focus on 11 specialties and regroup the 22 others into an aggregate called “Others.” In the most general model (Model 1), results from the first transition suggest that in all specialties where the estimated MC coefficient is significant (Thoracic Surgery, Gastroenterology, Obstetrics \& Gynecology, Pulmonology, Pediatrics, Others), the reform reduces the risk of discharge to home. This means that the reform increases the LOS in hospital for these specialties. The estimated MC coefficient is positive and significant only for Cardiology, Psychiatry, and Urology.

\begin{table}[htbp]
  \begin{center}
\addtolength{\leftskip} {-2.2cm}
  \caption{Effect (in log) of the reform on patients' risks of transition (by specialty)}
    \resizebox{1.3\textwidth}{!}{
    \begin{tabular}{lccccccccccr}
    \toprule \label{table_speciality}
          & \multicolumn{2}{c}{Model 1} & \multicolumn{2}{c}{Model 2} & \multicolumn{2}{c}{Model 3} & \multicolumn{2}{c}{Model 4} & \multicolumn{2}{c}{Model 5} & \multicolumn{1}{c}{Parallel trend} \\
    Variables & Estimate & SD    & Estimate & SD    & Estimate & SD    & Estimate & SD    & Estimate & SD    & \multicolumn{1}{c}{p-value} \\
    \midrule
    \textbf{1. Ad(RAd) → Home} &       &       &       &       &       &       &       &       &       &       &  \\
    MC: Cardiology & 0.038 & 0.014 & 0.031 & 0.010 & -0.020 & 0.013 & -0.008 & 0.013 & 0.013 & 0.013 & \multicolumn{1}{c}{0.957} \\
    MC: Gen. Surgery & -0.014 & 0.010 & -0.002 & 0.008 & -0.034 & 0.010 & -0.050 & 0.010 & -0.056 & 0.009 & \multicolumn{1}{c}{0.890} \\
    MC: Thor. Surgery & -0.068 & 0.032 & 0.004 & 0.025 & -0.095 & 0.032 & -0.085 & 0.032 & -0.041 & 0.027 & \multicolumn{1}{c}{0.874} \\
    MC: Gastro. & -0.144 & 0.029 & -0.188 & 0.022 & -0.171 & 0.029 & -0.225 & 0.029 & -0.143 & 0.027 & \multicolumn{1}{c}{0.587} \\
    MC: Obs. \& Gyn. & -0.036 & 0.011 & -0.056 & 0.007 & -0.059 & 0.010 & -0.069 & 0.010 & -0.009 & 0.009 & \multicolumn{1}{c}{0.758} \\
    MC: Pulmonology & -0.102 & 0.031 & -0.099 & 0.023 & -0.101 & 0.031 & -0.102 & 0.031 & -0.124 & 0.030 & \multicolumn{1}{c}{0.099} \\
    MC: Inter. Med. & -0.024 & 0.016 & -0.027 & 0.012 & -0.043 & 0.015 & -0.049 & 0.015 & -0.065 & 0.013 & \multicolumn{1}{c}{0.742} \\
    MC: ORL \& CFS & -0.031 & 0.026 & -0.061 & 0.019 & -0.042 & 0.026 & -0.067 & 0.026 & -0.038 & 0.024 & \multicolumn{1}{c}{0.199} \\
    MC: Pediatrics & -0.106 & 0.022 & -0.109 & 0.015 & -0.108 & 0.022 & -0.120 & 0.022 & -0.120 & 0.015 & \multicolumn{1}{c}{0.320} \\
    MC: Psychiatry & 0.090 & 0.024 & 0.123 & 0.018 & 0.078 & 0.023 & 0.077 & 0.023 & -0.113 & 0.015 & \multicolumn{1}{c}{0.581} \\
    MC: Urology & 0.056 & 0.027 & 0.044 & 0.020 & 0.092 & 0.027 & 0.081 & 0.027 & 0.155 & 0.026 & \multicolumn{1}{c}{0.731} \\
    MC: Others & -0.119 & 0.009 & -0.074 & 0.007 & -0.167 & 0.009 & -0.156 & 0.009 & -0.196 & 0.008 & \multicolumn{1}{c}{0.092} \\
    Female & -0.007 & 0.005 & -0.017 & 0.003 & -0.002 & 0.005 & 0.032 & 0.005 & -0.002 & 0.005 &  \\
    Age   & -0.003 & 0.000 & -0.005 & 0.000 & -0.001 & 0.000 & 0.000 & 0.000 & -0.001 & 0.000 &  \\
    (Age/150)$^2$ & -0.021 & 0.001 & -0.009 & 0.000 & -0.022 & 0.001 & -0.030 & 0.001 & -0.022 & 0.001 &  \\
    $\phi$ & 0.001 & 0.003 & -     & -     & 0.002 & 0.003 & -0.005 & 0.003 & 0.002 & 0.003 &  \\
    \midrule
    \textbf{2. Ad(RAd) → Death} &       &       &       &       &       &       &       &       &       &       &  \\
    MC: Cardiology & 0.024 & 0.058 & 0.031 & 0.057 & -0.042 & 0.054 & -0.096 & 0.054 & -0.052 & 0.052 & \multicolumn{1}{c}{0.659} \\
    MC: Gen. Surgery & 0.093 & 0.060 & 0.110 & 0.060 & 0.138 & 0.058 & 0.221 & 0.058 & 0.096 & 0.050 & \multicolumn{1}{c}{0.797} \\
    MC: Thor. Surgery & 0.109 & 0.155 & 0.190 & 0.152 & 0.030 & 0.153 & 0.018 & 0.153 & -0.122 & 0.129 & \multicolumn{1}{c}{0.126} \\
    MC: Gastro. & 0.177 & 0.120 & 0.163 & 0.117 & 0.186 & 0.119 & 0.309 & 0.118 & 0.122 & 0.108 & \multicolumn{1}{c}{0.698} \\
    MC: Obs. \& Gyn. & 0.570 & 0.223 & 0.605 & 0.222 & 0.619 & 0.224 & 0.817 & 0.220 & 0.553 & 0.174 & \multicolumn{1}{c}{NA} \\
    MC: Pulmonology & 0.001 & 0.088 & 0.035 & 0.083 & 0.076 & 0.087 & 0.134 & 0.086 & 0.072 & 0.085 & \multicolumn{1}{c}{0.991} \\
    MC: Inter. Med. & -0.077 & 0.046 & -0.094 & 0.043 & 0.070 & 0.042 & 0.116 & 0.041 & 0.112 & 0.037 & \multicolumn{1}{c}{0.376} \\
    MC: ENT \& CFS & 0.463 & 0.224 & 0.401 & 0.222 & 0.474 & 0.224 & 0.508 & 0.221 & 0.269 & 0.198 & \multicolumn{1}{c}{0.627} \\
    MC: Pediatrics & 0.486 & 0.335 & 0.500 & 0.335 & 0.535 & 0.330 & 0.537 & 0.332 & -0.082 & 0.211 & \multicolumn{1}{c}{NA} \\
    MC: Urology & -0.685 & 0.302 & -0.675 & 0.300 & -0.649 & 0.300 & -0.637 & 0.300 & -0.585 & 0.298 & \multicolumn{1}{c}{NA} \\
    MC: Others & -0.067 & 0.035 & -0.056 & 0.034 & -0.104 & 0.034 & -0.180 & 0.034 & -0.095 & 0.032 & \multicolumn{1}{c}{0.787} \\
    Female & -0.079 & 0.017 & -0.112 & 0.016 & -0.078 & 0.017 & -0.153 & 0.017 & -0.078 & 0.017 &  \\
    Age   & 0.023 & 0.004 & 0.016 & 0.003 & 0.024 & 0.003 & 0.052 & 0.003 & 0.024 & 0.003 &  \\
    (Age/150)$^2$ & 0.006 & 0.004 & 0.020 & 0.004 & 0.005 & 0.004 & -0.024 & 0.004 & 0.005 & 0.004 &  \\
    $\phi$ & -0.341 & 0.020 & -     & -     & -0.345 & 0.020 & -0.342 & 0.017 & -0.349 & 0.019 &  \\
    \midrule
    \textbf{3. Home → RAd} &       &       &       &       &       &       &       &       &       &       &  \\
    MC: Cardiology & -0.109 & 0.093 & -0.099 & 0.091 & -0.011 & 0.091 & -0.022 & 0.090 & -0.097 & 0.084 & \multicolumn{1}{c}{0.521} \\
    MC: Gen. Surgery & 0.407 & 0.084 & 0.415 & 0.082 & 0.481 & 0.082 & 0.528 & 0.081 & 0.384 & 0.066 & \multicolumn{1}{c}{0.949} \\
    MC: Thor. Surgery & 0.311 & 0.284 & 0.316 & 0.282 & 0.339 & 0.283 & 0.349 & 0.283 & 0.640 & 0.236 & \multicolumn{1}{c}{NA} \\
    MC: Gastro. & 0.009 & 0.131 & -0.046 & 0.125 & 0.118 & 0.130 & 0.214 & 0.129 & 0.283 & 0.118 & \multicolumn{1}{c}{0.728} \\
    MC: Obs. \& Gyn. & 0.637 & 0.096 & 0.644 & 0.095 & 0.729 & 0.093 & 0.755 & 0.093 & 0.396 & 0.071 & \multicolumn{1}{c}{0.658} \\
    MC: Pulmonology & -0.025 & 0.113 & 0.020 & 0.102 & 0.021 & 0.112 & 0.017 & 0.111 & 0.159 & 0.108 & \multicolumn{1}{c}{0.310} \\
    MC: Inter. Med. & 0.397 & 0.083 & 0.401 & 0.080 & 0.416 & 0.080 & 0.437 & 0.080 & 0.390 & 0.065 & \multicolumn{1}{c}{0.069} \\
    MC: ORL \& CFS & 0.262 & 0.217 & 0.256 & 0.213 & 0.363 & 0.216 & 0.427 & 0.216 & 0.450 & 0.193 & \multicolumn{1}{c}{0.476} \\
    MC: Pediatrics & 0.835 & 0.107 & 0.899 & 0.103 & 0.782 & 0.107 & 0.805 & 0.106 & 0.399 & 0.067 & \multicolumn{1}{c}{0.984} \\
    MC: Psychiatry & 0.513 & 0.080 & 0.482 & 0.075 & 0.442 & 0.077 & 0.436 & 0.077 & 0.245 & 0.048 & \multicolumn{1}{c}{0.639} \\
    MC: Urology & -0.314 & 0.201 & -0.314 & 0.197 & -0.255 & 0.200 & -0.217 & 0.199 & -0.187 & 0.197 & \multicolumn{1}{c}{0.729} \\
    MC: Others & -0.112 & 0.049 & -0.077 & 0.045 & -0.044 & 0.048 & -0.027 & 0.047 & 0.039 & 0.043 & \multicolumn{1}{c}{0.737} \\
    Female & 0.000 & 0.020 & 0.032 & 0.018 & -0.001 & 0.020 & -0.040 & 0.020 & 0.000 & 0.020 &  \\
    Age   & -0.010 & 0.002 & -0.008 & 0.002 & -0.016 & 0.002 & -0.013 & 0.002 & -0.016 & 0.002 &  \\
    (Age/150)$^2$ & 0.005 & 0.003 & -0.004 & 0.003 & 0.012 & 0.003 & 0.016 & 0.003 & 0.012 & 0.003 &  \\
    $\psi$ & -0.195 & 0.015 & -     & -     & -0.187 & 0.015 & -0.280 & 0.015 & -0.189 & 0.015 &  \\
    \bottomrule
    \end{tabular}%
    }
\end{center}
  
\end{table}%

\pagebreak

\begin{table}[!t]
  \begin{center}
 \addtolength{\leftskip} {-2.2cm}  
   (Continued) \\
    \resizebox{1.3\textwidth}{!}{
    \begin{tabular}{lccccccccccr}
    \toprule
          & \multicolumn{2}{c}{Model 1} & \multicolumn{2}{c}{Model 2} & \multicolumn{2}{c}{Model 3} & \multicolumn{2}{c}{Model 4} & \multicolumn{2}{c}{Model 5} & \multicolumn{1}{c}{Parallel trend} \\
    Variables & Estimate & SD    & Estimate & SD    & Estimate & SD    & Estimate & SD    & Estimate & SD    & \multicolumn{1}{c}{p-value} \\
    \midrule
    \textbf{4. Home → Death} &       &       &       &       &       &       &       &       &       &       &  \\
    MC: Cardiology & -0.061 & 0.064 & -0.047 & 0.060 & -0.054 & 0.060 & -0.086 & 0.059 & -0.061 & 0.058 & \multicolumn{1}{c}{0.780} \\
    MC: Gen. Surgery & 0.331 & 0.052 & 0.294 & 0.047 & 0.350 & 0.050 & 0.388 & 0.049 & 0.195 & 0.043 & \multicolumn{1}{c}{0.883} \\
    MC: Thor. Surgery & 0.245 & 0.149 & 0.209 & 0.138 & 0.139 & 0.147 & 0.143 & 0.145 & 0.133 & 0.128 & \multicolumn{1}{c}{0.509} \\
    MC: Gastro. & -0.037 & 0.114 & -0.054 & 0.102 & 0.028 & 0.112 & 0.219 & 0.110 & 0.070 & 0.104 & \multicolumn{1}{c}{0.737} \\
    MC: Obs. \& Gyn. & 0.204 & 0.140 & 0.209 & 0.131 & 0.237 & 0.140 & 0.517 & 0.135 & 0.297 & 0.113 & \multicolumn{1}{c}{0.391} \\
    MC: Pulmonology & -0.153 & 0.093 & -0.130 & 0.079 & -0.110 & 0.092 & -0.043 & 0.090 & -0.076 & 0.090 & \multicolumn{1}{c}{0.738} \\
    MC: Inter. Med. & 0.188 & 0.053 & 0.165 & 0.048 & 0.174 & 0.050 & 0.225 & 0.050 & 0.171 & 0.044 & \multicolumn{1}{c}{0.447} \\
    MC: ORL \& CFS & 0.046 & 0.143 & 0.123 & 0.129 & 0.031 & 0.142 & 0.022 & 0.139 & 0.062 & 0.131 & \multicolumn{1}{c}{0.482} \\
    MC: Psychiatry & 0.339 & 0.181 & 0.306 & 0.176 & 0.293 & 0.178 & 0.303 & 0.177 & 0.152 & 0.114 & \multicolumn{1}{c}{NA} \\
    MC: Urology & 0.233 & 0.119 & 0.218 & 0.108 & 0.204 & 0.116 & 0.212 & 0.115 & 0.210 & 0.114 & \multicolumn{1}{c}{0.375} \\
    MC: Others & -0.064 & 0.034 & -0.065 & 0.030 & -0.063 & 0.032 & -0.141 & 0.032 & -0.027 & 0.030 & \multicolumn{1}{c}{0.421} \\
    Female & -0.220 & 0.017 & -0.172 & 0.014 & -0.220 & 0.017 & -0.328 & 0.017 & -0.219 & 0.017 &  \\
    Age   & 0.004 & 0.003 & 0.012 & 0.003 & 0.008 & 0.003 & 0.032 & 0.003 & 0.008 & 0.003 &  \\
    (Age/150)$^2$ & 0.053 & 0.004 & 0.035 & 0.004 & 0.050 & 0.004 & 0.036 & 0.004 & 0.050 & 0.004 &  \\
    $\psi$ & -0.613 & 0.016 & -     & -     & -0.603 & 0.016 & -0.797 & 0.016 & -0.603 & 0.016 &  \\
    \midrule
    \textit{log(likelihood)} & \multicolumn{2}{c}{-2345239} & \multicolumn{2}{c}{-2301265} & \multicolumn{2}{c}{-2349294} & \multicolumn{2}{c}{-2367719} & \multicolumn{2}{c}{-2350000} &  \\
    Number of patients & \multicolumn{2}{c}{320,441} & \multicolumn{2}{c}{320,441} & \multicolumn{2}{c}{320,441} & \multicolumn{2}{c}{320,441} & \multicolumn{2}{c}{320,441} &  \\
    Number of observations & \multicolumn{2}{c}{1,400,121} & \multicolumn{2}{c}{1,400,121} & \multicolumn{2}{c}{1,400,121} & \multicolumn{2}{c}{1,400,121} & \multicolumn{2}{c}{1,400,121} &  \\
    \midrule
    Unobserved heterogeneity & \multicolumn{2}{c}{Yes} & \multicolumn{2}{c}{No} & \multicolumn{2}{c}{Yes} & \multicolumn{2}{c}{Yes} & \multicolumn{2}{c}{Yes} &  \\
    Hospital FE & \multicolumn{2}{c}{80 largest} & \multicolumn{2}{c}{80 largest} & \multicolumn{2}{c}{10 largest} & \multicolumn{2}{c}{10 largest} & \multicolumn{2}{c}{10 largest} &  \\
    Charlson co-morbidity index & \multicolumn{2}{c}{Yes} & \multicolumn{2}{c}{Yes} & \multicolumn{2}{c}{Yes} & \multicolumn{2}{c}{No} & \multicolumn{2}{c}{Yes} &  \\
    Department size (number of specialists) & \multicolumn{2}{c}{Yes} & \multicolumn{2}{c}{Yes} & \multicolumn{2}{c}{Yes} & \multicolumn{2}{c}{Yes} & \multicolumn{2}{c}{Yes} &  \\
    12 Specialty FE & \multicolumn{2}{c}{Yes} & \multicolumn{2}{c}{Yes} & \multicolumn{2}{c}{Yes} & \multicolumn{2}{c}{Yes} & \multicolumn{2}{c}{Yes} &  \\
    18 Diagnoses FE & \multicolumn{2}{c}{Yes} & \multicolumn{2}{c}{Yes} & \multicolumn{2}{c}{Yes} & \multicolumn{2}{c}{Yes} & \multicolumn{2}{c}{Yes} &  \\
    19 Region FE & \multicolumn{2}{c}{Yes} & \multicolumn{2}{c}{Yes} & \multicolumn{2}{c}{Yes} & \multicolumn{2}{c}{Yes} & \multicolumn{2}{c}{Yes} &  \\
    Linear \& quadratic trends & \multicolumn{2}{c}{Yes} & \multicolumn{2}{c}{Yes} & \multicolumn{2}{c}{Yes} & \multicolumn{2}{c}{Yes} & \multicolumn{2}{c}{No} &  \\
    Year FE & \multicolumn{2}{c}{Yes} & \multicolumn{2}{c}{Yes} & \multicolumn{2}{c}{Yes} & \multicolumn{2}{c}{Yes} & \multicolumn{2}{c}{Yes} &  \\    \bottomrule
    \end{tabular}%
    }
\end{center}

\scriptsize
\textit{Note:} All specifications use a multi-state multi-spell proportional conditional hazard model. The baseline hazards are piecewise constant. In all models (except Model 2), unobserved heterogeneity is modelled using a mixed weighting of the values taken by \textit{iid} random variables drawn from a standard normal distribution. The parallel trend p-values correspond to Model 3. All the models (except Model 2) are estimated using simulated Maximum Likelihood. Model 2 does not account for unobserved heterogeneity and is estimated using Maximum Likelihood.  We only estimate the MC effect for specialties of interest that have a sufficient number of observations. Pediatrics does not appear in the fourth transition because we only observe 14 deaths at home under FFS. Pediatrics is included in the category of "Others" in this case for this transition. Similarly, Psychiatry is also included in the "Others " category for the third transition, given that we have only eight observations under FFS.

\end{table}%

With regard to the second transition (from hospital to death), the MC coefficient is significant for very few specialties, as expected from the results of Table \ref{table_overall}. For the latter specialties, it is positive for Obstetrics \& Gynecology, and ORL. It is negative only for Urology. In short, the reform seems to have little effect on the quality of specialists' healthcare services, at least as measured by this quality indicator.

With regard to the results of the third transition, which provides an additional quality indicator, they indicate that in all specialties where the MC coefficient is significant (\textit{i.e.}, General Surgery, Obstetric \& Gynecology, Internal Medicine, Pediatrics, Psychiatry), the reform increases the risk of re-admission within thirty days after discharge. Furthermore, the reform's effect on the “Others” category is negative but small in terms of its absolute value. Recall that at the aggregate level, results from Table 4 suggest that the effect of the MC reform is to increase the risk of re-admission to hospital within thirty days after discharge.

The MC coefficient in the fourth transition represents the quality indicator providing the reform's effect on the risk of dying within one year after discharge for each specialty. The coefficient is positive for all specialties for which it is significant: General Surgery, Internal Medicine, 
and Urology.  This third indicator further suggests that the reform reduced the quality of specialists' healthcare services. Lastly, results regarding the parallel trend test are not rejected for all specialties and transitions.\footnote{The term NA in Table \ref{table_speciality} indicates that the test could not be run for some specialties because the data set was too small.}  

\subsection{Dynamic analysis}
This section extends Model 1 in Table \ref{table_overall} to analyze some dynamic elements of the reform's effect. To reach this objective, we allow the MC coefficient (= ATT) to vary across the intervals of years that physicians who adopt MC would spend under this compensation system (denoted Exp). In particular, we study the reform's effect on the risk of the four transitions depending on how long the department of the treating physician has been under MC.\

Regarding the first transition, results from Table \ref{Short_long} suggest that the patient's risk of being discharged to home decreases more and more as his treating physician, through his department, is under MC for a long time. When the physician has worked under MC for less than two years, the effect is -4.6\% (and is significant), while the effect is -15.1\% (and is significant) when the physician has been under MC for over 10 years. However, the effect of MC is never significant (given large SDs relative to the means) on the hospitalized patient's risk of dying (second transition), regardless of whether their treating physician's department is an early or later MC adopter.\

The reform's effect on the risk of hospital re-admission (third transition) is positive and significant when the treating physician is under MC for two years and more. The effect is increasing, i.e., 0.247 when the physician spends between two and four years under MC and attaining a value of 0.400 when the physician has accumulated ten years or more under MC.\

Finally, regarding the reform's effect on the patient's risk of dying within one year after discharge, it is positive and increases when the treating physician is under MC for more than five years. The effect is 7.5\% when the physician is under MC for more than five years but less than nine years; it reaches 21\% when the physician is under MC for ten years or more. \ 

In short, based on transition 3 and 4 results, the reform's effect on the quality of MC physicians' healthcare services is negative. Moreover, this effect increases (in absolute value), given that these physicians have been under MC for a long time. This may partly be explained by behavioural adjustments to the reform that will likely require more time.

\begin{table}[htbp]
\begin{center}
\caption{Short- \textit{vs} long-term effects (in log)}
    \resizebox{0.7\textwidth}{!}{
\begin{tabular}{lcc}
     \toprule \label{Short_long}
    Variables & Estimate & SD \\
     
    \midrule
    1. \textbf{Ad(RAd) → Home} &       &  \\
    Exp: < 2 years & -0.046 & 0.007 \\
    Exp: 2–4 years & -0.054 & 0.007 \\
    Exp: 5–9 years & -0.095 & 0.006 \\
    Exp: $\geq$ 10 years & -0.151 & 0.007 \\
    $\phi$ & 0.002 & 0.003 \\
    \midrule
    2. \textbf{Ad(RAd) → Death} &       &  \\
    Exp: < 2 years & -0.058 & 0.039 \\
    Exp: 2–4 years & 0.007 & 0.034 \\
    Exp: 5–9 years & 0.037 & 0.030 \\
    Exp: $\geq$ 10 years & 0.061 & 0.034 \\
    $\phi$ & -0.344 & 0.020 \\
    \midrule
    3. \textbf{Home → RAd} &       &  \\
    Exp: < 2 years & 0.047 & 0.042 \\
    Exp: 2–4 years & 0.247 & 0.037 \\
    Exp: 5–9 years & 0.224 & 0.036 \\
    Exp: $\geq$ 10 years & 0.400 & 0.041 \\
    $\psi$ & -0.188 & 0.015 \\
    \midrule
    4. \textbf{Home → Death} &       &  \\
    Exp: < 2 years & -0.025 & 0.038 \\
    Exp: 2–4 years & 0.021 & 0.033 \\
    Exp: 5–9 years & 0.075 & 0.029 \\
    Exp: $\geq$ 10 years & 0.212 & 0.033 \\
    $\psi$ & -0.597 & 0.016 \\
    \midrule
    \textit{log(likelihood)} & \multicolumn{2}{c}{-2350046} \\
    Number of patients & \multicolumn{2}{c}{320,441}  \\
    Number of observations & \multicolumn{2}{c}{1,400,121} \\
 \midrule
Unobserved heterogeneity & \multicolumn{2}{c}{Yes} \\
    Hospital FE & \multicolumn{2}{c}{80 largest}\\
    Charlson co-morbidity index & \multicolumn{2}{c}{Yes}  \\
    Department size (number of specialists) & \multicolumn{2}{c}{Yes}  \\
    12 Specialty FE & \multicolumn{2}{c}{Yes}   \\
    18 Diagnoses FE & \multicolumn{2}{c}{Yes} \\
    19 Region FE & \multicolumn{2}{c}{Yes}  \\
    Linear \& quadratic trends & \multicolumn{2}{c}{Yes}   \\
    Year FE & \multicolumn{2}{c}{Yes}  \\   
    \bottomrule
    \end{tabular}  
    }
\end{center}
\scriptsize

\textit{Note:} This specification, corresponding to Model 1, uses an MSMS proportional conditional hazard model. The baseline hazards are piecewise constant. Unobserved heterogeneity is modelled using a mixed weighting of the values taken by \textit{iid} random variables drawn from a standard normal distribution.  The model is estimated using simulated Maximum Likelihood.  
\end{table}

\begin{table}[htbp]
\begin{center}
  \caption{Unconditional Effect of the Reform on Durations  (ATT in days)}
    \resizebox{0.7\textwidth}{!}{
    \begin{tabular}{lcccc}
    \toprule \label{tau_duration}
          & \multicolumn{2}{c}{Model 3} & \multicolumn{2}{c}{Model 3} \\
          & \multicolumn{2}{c}{(with heterogeneity)} & \multicolumn{2}{c}{(without heterogeneity)} \\
    \midrule
          & Estimate & Std. Err & Estimate & Std. Err \\
    \midrule
    1. \textbf{Ad(RAd) → Home} &       &       &       &  \\
    Overall & 0.75  & 0.04  & 0.84  & 0.04 \\
    Cardiology & 0.11  & 0.06  & 0.12  & 0.07 \\
    Gen. Surgery & 0.20  & 0.06  & 0.16  & 0.06 \\
    Thor. Surgery & 0.85  & 0.31  & 0.34  & 0.29 \\
    Gastro. & 1.53  & 0.25  & 2.38  & 0.27 \\
    Obs. \& Gyn. & 0.17  & 0.03  & 0.28  & 0.02 \\
    Pulmonology & 0.80  & 0.24  & 1.11  & 0.24 \\
    Inter. Med. & 0.48  & 0.16  & 0.79  & 0.18 \\
    ORL \& CFS & 0.14  & 0.09  & 0.33  & 0.08 \\
    Pediatrics & 0.35  & 0.07  & 0.42  & 0.06 \\
    Psychiatry & -2.06 & 0.67  & -4.15 & 0.61 \\
    Urology & -0.32 & 0.09  & -0.28 & 0.09 \\
    Others & 1.80  & 0.08  & 1.66  & 0.09 \\
    \midrule
   2. \textbf{Ad(RAd) → Death} &       &       &       &  \\
    Overall & -0.20 & 0.28  & -0.25 & 0.26 \\
    Cardiology & 0.32  & 0.44  & 0.32  & 0.42 \\
    Gen. Surgery & -1.87 & 0.89  & -2.00 & 0.79 \\
    Thor. Surgery & -0.14 & 0.80  & -0.49 & 0.72 \\
    Gastro. & -2.37 & 1.47  & -2.35 & 1.47 \\
    Obs. \& Gyn. & -10.31 & 3.78  & -10.91 & 3.79 \\
    Pulmonology & -0.39 & 0.44  & -0.59 & 0.44 \\
    Inter. Med. & -0.76 & 0.42  & -0.29 & 0.45 \\
    ORL \& CFS & -12.27 & 5.57  & -10.18 & 5.90 \\
    Pediatrics & -10.66 & 5.94  & -11.00 & 6.51 \\
    Urology & 6.81  & 3.38  & 6.70  & 3.21 \\
    Others & 1.48  & 0.43  & 1.08  & 0.48 \\
    \midrule
    3. \textbf{Home → RAd} &       &       &       &  \\
    Overall & -2.98 & 0.34  & -3.18 & 0.33 \\
    Cardiology & 0.14  & 1.33  & -0.29 & 1.19 \\
    Gen. Surgery & -6.02 & 1.00  & -6.21 & 1.02 \\
    Thor. Surgery & -3.38 & 2.81  & -3.52 & 2.85 \\
    Gastro. & -1.85 & 1.71  & -0.78 & 1.95 \\
    Obs. \& Gyn. & -8.99 & 1.19  & -9.21 & 1.12 \\
    Pulmonology & -0.33 & 1.76  & -1.38 & 1.63 \\
    Inter. Med. & -5.41 & 0.99  & -5.33 & 1.00 \\
    ORL \& CFS & -5.18 & 3.24  & -5.56 & 3.00 \\
    Pediatrics & -9.71 & 1.45  & -10.73 & 1.25 \\
    Psychiatry & -5.71 & 0.98  & -5.14 & 0.93 \\
    Urology & 4.16  & 3.04  & 3.73  & 3.23 \\
    Others & 0.68  & 0.71  & -0.16 & 0.70 \\
    \midrule
    4. \textbf{Home → Death} &       &       &       &  \\
    Overall & -7.45 & 2.55  & -5.55 & 2.07 \\
    Cardiology & 6.05  & 7.55  & 5.04  & 6.27 \\
    Gen. Surgery & -43.91 & 6.50  & -38.60 & 5.65 \\
    Thor. Surgery & -15.64 & 16.19 & -11.44 & 15.28 \\
    Gastro. & -2.80 & 10.54 & 1.94  & 10.11 \\
    Obs. \& Gyn. & -25.42 & 13.12 & -24.01 & 14.12 \\
    Pulmonology & 10.53 & 7.57  & 8.94  & 7.74 \\
    Inter. Med. & -16.35 & 4.90  & -14.63 & 4.20 \\
    ORL \& CFS & -3.75 & 17.38 & -13.16 & 15.86 \\
    Psychiatry & -35.95 & 20.84 & -30.67 & 21.31 \\
    Urology & -28.70 & 14.58 & -24.25 & 14.82 \\
    Others & 6.64  & 3.50  & 7.56  & 3.02 \\
    \bottomrule
    \end{tabular}%
  \label{tab:addlabel}%
    }
\end{center}
\end{table}%

\subsection{The effect of the MC reform on duration}

Thus far, our findings have primarily focused upon the ATT of the reform on the patient's risk of each transition (in log) for all selected specialties. However, this doesn't offer a clear insight into how the reform impacts the patient's expected unconditional duration of stay in a given state (with transition $r$). Such information is crucial for policymakers and hospital managers. This is particularly evident for the duration of hospitalization, as it directly correlates with resource allocation and costs. Similarly, the duration before rehospitalization after discharge not only pertains to costs but also to effective resource management. For instance, a higher frequency of patients readmitted within a week of discharge suggests that a greater emphasis might be needed on post-discharge care or follow-up interventions during the initial days after leaving the hospital. Furthermore, the duration before death post-discharge provides distinct viewpoints on patient outcomes and potential costs. A shorter duration between discharge and death might indicate premature discharges or insufficient post-discharge support, leading to potentially avoidable costs linked to end-of-life care outside the hospital.

Appendix A provides the formal relationship between an  unconditional duration expectation in a stay and the conditional hazard function. More heuristically, to simulate the ATT of the reform on a duration expectation, one would first calculate the density function of the duration, given that there is a formula that links the hazard to the density. The hazard is nothing more than the density divided by the expression one minus the distribution function. Note that for the case with heterogeneity, we need to integrate the density with respect to heterogeneity to obtain an unconditional density. The unconditional duration expectation is then obtained from integration in the same way that we would calculate an expectation by knowing the density function. Finally, we differentiate between the duration expectations when the treating physician is working in a department under MC minus the duration expectations when the department is under FFS to compute the ATT of the reform on the patient's duration expectations.

Table \ref{tau_duration} provides results for unconditional ATT on the expected duration of stay (in days) for each state, overall specialties, and selected specialties. We use Model 3 (10 hospitals) rather than Model 1 (80 hospitals), given that the latter is non-tractable because of the large number of integrations that are required to perform in the model with heterogeneity and that many standard errors cannot be computed in the model without heterogeneity. For the sake of comparison, we also present Model 3 without heterogeneity.

When the patient is in hospital (with the transition to Home), being in a department under MC increases their duration at the hospital by 0.75 days on average (= ATT), and the effect is significant. This is consistent with the negative effect of the reform on the hazard of leaving hospital to go home (see Table \ref{table_overall}). The ATT is also significant and positive for almost all specialties (except Psychiatry and Urology, for which it is negative and significant). The five specialties for which the ATT is the largest are Others (1.80 days), Gastroenterology (1.53 days), Thoracic Surgery (0.85 days), Pulmonology  (0.80 days) and Internal Medicine (0.35 days). The model's results without heterogeneity follow the same pattern (same sign) though the size may differ. For instance, the ATT for Gastroenterology is much higher (2.38 days).

Regarding the duration in hospital when the transition is to death, our results are also consistent with our expectations, indicating that the effect (unconditional ATT) of the reform is non-significant on average. The duration before death decreases, and this effect is significant for ORL \& Cervicofacial Surgery (-12.27), for Obstetrics \& Gynecology (-10.31), and  General Surgery (-1.87).  In contrast, the duration before death significantly increases for Urology (6.81) and Others (1.48). As was the case for the ATT with the transition to Home, the results are quite similar in the model without heterogeneity.

The reform also affects the duration of stay at home after discharge and before re-admission to hospital with the same diagnosis. An important indicator of the effect of MC on the quality of healthcare services that are provided by treating specialists can be measured by its effect on this duration. According to our results, the reform decreased duration by 2.98 days on average before patient re-admission when an MC specialist treated a patient. When significant, the ATT is negative for all selected specialties. The five specialties for which the ATT is the largest in terms of absolute value are Pediatrics (-9.71), Obstetrics \& Gynecology (-8.99), General Surgery (-6.02), Psychiatry (-5.71), and Internal Medicine (-5.41). This again suggests that based on this quality indicator, the reform's effect negatively affected the quality of healthcare services by specialists. Note that the corresponding results are very close when based upon the model without heterogeneity. 

Finally, the ATT of the reform on the duration of the period after discharge to death provides an additional measure of its effect on the quality of healthcare services by MC specialists. Our results suggest that the reform reduced by 7.45 days the duration of the period after discharge to death. When significant, the ATT is negative for each specialty. The specialties for which the ATT is largest (in absolute value) are 
General Surgery (-43.91), Urology (-28.70) and Internal Medicine (-16.35). Based on this indicator and in conformity with our hazard results, these results indicate that the MC reform negatively affects the quality of healthcare services by MC specialists.

\section{Discussions and Conclusion}

Our paper is one of the first to evaluate the effect of a physician's mixed compensation system on the quality of their healthcare services based upon a natural experiment. This system was introduced in Quebec in 1999 and blended salary (\textit{per diem}) and partial fee per clinical service. We develop an MSMS hazard model with correlated heterogeneity analogous to a difference-in-differences approach.  Our results indicate that the reform reduced the quality of healthcare services provided by MC specialists. Our results suggest that the risk of re-hospitalization of their patients within 30 days after discharge increased by 17.8\%, and the risk of dying within one year after discharge increased by 6.2\%. As a result, the average spell at home before re-hospitalization decreased by three days, and the average spell at home before death after discharge decreased by 7.45 days. These results could be attributed to the fact that being partly paid by salary (\textit{per diem}), independent of the quantity and quality of clinical services, may encourage some MC specialists to evade medical practice for alternative allocation of time in hospital, which is unpaid under the FFS system (such as administrative and teaching duties) and to reduce their effort when performing clinical services.

Our results show that the reform increased the LOS in hospital by 0.75 days (\textit{i.e.}, a 5.7\% decrease in the probability of being discharged) for a patient treated by a MC specialist. Yet, the effect of MC on LOS is less directly interpretable in terms of quality. Referring to a time-based approach, one could consider that the increase in the duration of hospitalization may have had beneficial effects on patients' health insofar as the MC doctor can take longer to cure the patient \citep[\textit{e.g.,}][]{carey2015, heggestad2002}. However, we can also consider the opposite effect: hospitalizations that are too long can be considered inappropriate because they can lead to complications for hospitalized patients and may be ineffective given the costs that they incur \citep[\textit{e.g.,}][]{regenbogen2017}. The length of hospitalization directly correlates with complications following treatment, whether it is a simple procedure such as cataract surgery or a more complicated procedure such as heart-bypass surgery. If the treatment is badly performed or is not of good quality, the duration of hospitalization will increase, given the complications that it causes.

The MC reform has brought some positive effects to the healthcare system. In particular, one expects that it has reduced the number of unnecessary medical services to some degree. Previous studies \citep[\textit{e.g.}, see][]{dumont2008,fortin2021} indicate that the volume of clinical services significantly decreased following the reform. Yet, no evidence provides information regarding the proportion of this reduction that can be judged unnecessary. The reform has also allowed MC doctors with more complex medical activities (pediatricians, psychiatrists, surgeons, \textit{etc.}) to spend more time with their patients, promoting teamwork and, consequently, their patients' health outcomes. Also, the reform may have encouraged some MC doctors to devote more time to training their students during the \textit{per diems}, thus making them more competent in providing healthcare. Besides, our results show that the reform's effect does not correspond to some expected objectives, particularly regarding the quality of medical services measured by output-based indicators.

Of course, our analysis does not apply to all MC systems. One basic advantage of the MC payment mechanism is that, by definition, it can use more instruments than a simple FFS, salary or capitation system \citep{MaMcguire1997}. Basic public economics theory suggests that a policymaker needs at least the same number of instruments as the objectives he wants to achieve (here, the quality and quantity of medical services). Therefore, the MC payment system is a natural approach to use for this matter. Yet, it may be hard to determine the optimal level of each instrument. In particular, an appropriate MC system could be developed so that its parameters vary across specialties and regions (to improve the spatial equity in allocating medical services, for instance) or according to patients' health problem complexity. Determining a physician MC mechanism that is optimal for patients, doctors, and society, in general, is a crucial issue that needs more research. Developing and estimating an econometric structural approach could be most useful in this regard. The analysis should also consider the cost of implementing such a system, given its potential complexity.

\bibliographystyle{chicago}
\bibliography{Referencesup}
\newpage

\appendix
\section{ATT on the log of the hazard and duration}
As pointed out in Section \ref{DiD}, the average treatment effects on the treated (ATT) on the logarithm of the conditional hazard for the transition $r \in \mathcal{T}$ is the parameter $\theta_{(r)}$. Indeed, from Equations \eqref{risque} and \eqref{eq:xbI}, the log of the conditional hazard for the departments that have adopted MC is given by 
\begin{equation}
    \log\big(\lambda_{(r)}^{I}(t, MC)\big) = \log(\lambda_{(r)0}(t)) + \theta_{(r)}MC(T) + z_{(r)}(T)^N + \log(\nu_{(r)}). \label{eq:LI}
\end{equation}
\noindent The ATT on the log of the hazard is then $\log\big(\lambda_{(r)}^{I}(t, MC = 1)\big) - \log\big(\lambda_{(r)}^{I}(t, MC = 0)\big) = \theta_{(r)}$. The advantage of this measure lies in its simplicity. It depends neither upon the unobserved patient heterogeneity $\nu_{(r)}$, nor upon the control variables included in $z_{(r)}(T)^N$ (see Equation \eqref{eq:xbN}). Importantly, it does not depend on whether we condition or not on $\nu_{(r)}$.

Computing the ATT on the duration is more challenging as the model becomes nonlinear \citep[see][]{athey_2006} and requires numerical integrations. Moreover, the unobserved heterogeneity term $\nu_{(r)}$ does not cancel out, as is the case of the log of the hazard. Therefore, we compute the ATT of the expected duration, where the expectation is taken with respect to the unobserved patient heterogeneity $\nu_{(r)}$ and to the control variable (i.e., $z_{(r)}(T)^N$). 

Our approach to computing the unconditional ATT on the duration can be described in several steps. 
\paragraph{Step 1.} From Equation \eqref{eq:LI}, we first compute the survival conditionally on $\nu_{(r)}$ and $z_{(r)}(T)^N$. The survival function is given by
$$
    S_{(r)}^{I}(t, MC) = \exp\Big\{-\exp\big\{\theta_{(r)}MC(T) + z_{(r)}(T)^N\big\}\nu_{(r)}\int_0^t \lambda_{(r)0}(\tau)d\tau\Big\}.
$$
\noindent Note that $\int_0^t\lambda_{(r)0}(\tau)d\tau$ has a closed form because $\lambda_{(r)0}(\tau)$ is a step function. In particular, if the baseline hazard equates $\alpha_k$ on the interval $[\alpha_k^{-}, ~\alpha_k^+)$,  for $k= 1, 2, \dots$,  then $\int_0^t\lambda_{(r)0}(\tau)d\tau = 
\sum_{s = 1}^{k} \alpha_s(\min\{t, \alpha_s^+\}  - \alpha_s^-)$ for $t\in [\alpha_k^{-}, ~\alpha_k^+)$. 
\paragraph{Step 2.} From the survival function $S_{(r)}^{I}(t, MC)$, we compute the density function of the duration conditionally on $\nu_{(r)}$ and $z_{(r)}(T)^N$. This density function is defined by
\begin{align*}
    g_{(r)}^I(t, MC) &= -\dfrac{\partial S_{(r)}^{I}(t, MC)}{\partial t}.\\
    g_{(r)}^I(t, MC) &= \exp\big\{\theta_{(r)}MC(T) + z_{(r)}(T)^N\big\}\nu_{(r)} \lambda_{(r)0}(t) S_{(r)}^{I}(t, MC).
\end{align*}

\paragraph{Step 3.} We now take the expectation of $g_{(r)}^I(t, MC)$ to obtain the unconditional density function denoted $g_{(r)}^{I*}(t, MC)$, where the expectation is with respect to $\nu_{(r)}$ and $z_{(r)}(T)^N$. 
Unfortunately, $g_{(r)}^{I*}(t, MC)$ does not have a closed form. 

Given that $g_{(r)}^{I*}(t, MC)$ is an expectation with respect to two variables, we use the law of iterated expectations and approximate it using a two-stage approach.\footnote{For the model without unobserved patient heterogeneity, we directly compute the expectation with respect to $z_{(r)}(T)^N$.} In the first stage, we compute the expectation of $g_{(r)}^I(t, MC)$ with respect to the first variable $\nu_{(r)}$, and conditionally on the second variable $z_{(r)}(T)^N$. Let $\hat g_{(r)}^{I}(t, MC)$ be this expectation. Since $z_{(r)}(T)^N$ and $\nu_{(r)}$ are independent, $\hat g_{(r)}^{I}(t, MC)$ is approximated using simulations from the distribution of $\nu_{(r)}$, as is done for the log-likelihood \eqref{eq:hatloL}.

In the second stage, we compute $g_{(r)}^{I*}(t, MC)$, which is the expectation of $\hat g_{(r)}^{I}(t, MC)$ with respect to $z_{(r)}(T)^N$. We approximate this second expectation using a sample mean. Indeed, for each observation under the MC, we can approximate 
 $z_{(r)}(T)^N$ by replacing the parameters in Equation \eqref{eq:xbN} with their estimates. Then, $g_{(r)}^{I*}(t, MC)$ can be approximated using the mean of $\hat g_{(r)}^{I}(t, MC)$ in the subsample under the MC.

\paragraph{Step 4.}  From the unconditional density function $g_{(r)}^{I*}(t, MC)$, we obtain the expected duration given by
$$D^I(MC) = \int_0^{\infty} \tau g_{(r)}^{I*}(\tau, MC) d\tau.$$

\paragraph{Step 5.}  Finally, the ATT on the expected duration is $D^I(MC = 1) - D^I(MC = 0)$. We also compute the standard deviation of this ATT using the simulation approach proposed by
\cite{krinsky1990}.

\end{document}